\documentclass[twocolumn,tighten]{aastex61}

\usepackage{amsmath}
\usepackage{graphicx}
\usepackage{soul}
\usepackage{color}

\usepackage{amsmath} 

\newcommand{\gyr}{{\rm{Gyr}}}
\newcommand{\cmc}{{\tt{CMC}}}
\newcommand{\sse}{{\tt SSE}}
\newcommand{\bse}{{\tt BSE}}
\newcommand{\msun}{{\rm{M_\odot}}}

\newcommand{\pc}{{\rm{pc}}}

\newcommand{\kms}{{\rm{km\,s^{-1}}}}

\usepackage{hyperref}
\usepackage{longtable}

\usepackage{filecontents}

\begin{document}
\bibliographystyle{apj}

\title{Accreting black hole binaries in globular clusters}

\author[0000-0002-4086-3180]{Kyle Kremer}
\affil{\centering Department of Physics \& Astronomy, Northwestern University, Evanston, IL 60202, USA}
\affil{\centering Center for Interdisciplinary Exploration \& Research in Astrophysics (CIERA), Evanston, IL 60202, USA}
\email{kremer@u.northwestern.edu}
\author[0000-0002-3680-2684]{Sourav Chatterjee}
\affil{\centering Department of Physics \& Astronomy, Northwestern University, Evanston, IL 60202, USA}
\affil{\centering Center for Interdisciplinary Exploration \& Research in Astrophysics (CIERA), Evanston, IL 60202, USA}
\email{chatterjee.sourav2010@gmail.com}
\author{Carl L. Rodriguez}
\affil{\centering MIT-Kavli Institute for Astrophysics and Space Research, Cambridge, MA 02139, USA}
\author{Frederic A. Rasio}
\affil{\centering Department of Physics \& Astronomy, Northwestern University, Evanston, IL 60202, USA}
\affil{\centering Center for Interdisciplinary Exploration \& Research in Astrophysics (CIERA), Evanston, IL 60202, USA}

\begin{abstract}
We explore the formation of mass-transferring binary systems containing black holes within globular clusters. We show that it is possible to form mass-transferring black hole binaries with main sequence, giant, and white dwarf companions with a variety of orbital parameters in globular clusters spanning a large range in present-day properties. All mass-transferring black hole binaries found in our models at late times are dynamically created. The black holes in these systems experienced a median of $\sim 30$ dynamical encounters within the cluster before and after acquiring the donor. Furthermore, we show that the presence of mass-transferring black hole systems has little correlation with the total number of black holes within the cluster at any time. This is because the net rate of formation of black hole--non-black hole binaries in a cluster is largely independent of the total number of retained black holes. Our results suggest that the detection of a mass-transferring black hole binary in a globular cluster does not necessarily indicate that the host cluster contains a large black hole population.
\end{abstract}

\keywords{scattering -- methods: numerical -- stars: black holes -- stars: kinematics and dynamics -- globular clusters: general -- X-rays: binaries}

\section{Introduction} \label{sec:intro}
Dense star clusters are expected to form a large number of black holes (BHs) simply because of 
the large number ($N$) of stars they are born with and the properties of any reasonable initial stellar mass function (IMF). What happens to these BHs later on as a result of complex dynamical processing inside 
these clusters has been long debated and the understanding has evolved significantly over the past several decades. \citet{Spitzer1967} argued that BHs, being significantly more massive than typical 
stars in the cluster, would quickly mass segregate on sub-Gyr timescales, forming a compact sub-cluster that is 
dynamically decoupled from the rest of the cluster. It was argued that because of 
the compactness and low effective $N$ of the sub-cluster, the BHs will be ejected from the cluster as the result of 
mutual strong encounters on few-Gyr timescales. Thus, the old ($\gtrsim 10\,\gyr$) 
present-day globular clusters (GCs) were expected to retain, at most, a couple of BHs. Several rate-based theoretical studies supported this expectation \citep[e.g.,][]{Kulkarni1993, Sigurdsson1993, PortegiesZwart2000, Kalogera2004}. Furthermore, from an observational perspective, {\em all} X-ray binaries (XRBs) in GCs discovered prior to 2007 were found to have neutron star accretors bolstering the above expectation \citep[e.g.,][]{Zyl2004, Lewin2006, Altamirano2010, Altamirano2012,Bozzo2011}.

This classical understanding started to change as mass-transferring BH binary candidates began to be discovered in GCs, 
first in NGC\ 4472 \citep{Maccarone2007,Irwin2010} through 
high X-ray luminosity and high variability, and then in several GCs in the Milky Way 
(MW) by their relative X-ray and radio luminosities 
\citep{Strader2012,Chomiuk2013,Miller-Jones2014}. 

Modern, realistic numerical simulations also show that the classical argument of 
quick BH evaporation is not correct. In fact, the BH sub-cluster does not remain 
dynamically decoupled from the rest of the cluster for long periods of time, thus the 
actual timescale for evaporation of BHs is significantly longer
\citep[e.g.,][]{Mackay2008,Morscher2013,Morscher2015,Chatterjee2017a}.

The dynamical processing of BHs in a cluster and the effect of BH dynamics on 
the overall evolution of the host cluster are of high current interest, especially since 
the ground-breaking recent discoveries of merging binary BHs (BBHs) by LIGO \citep{Abbott2016a,Abbott2016b,Abbott2016c,Abbott2016d,Abbott2016e, Abbott2017}. In particular, 
it is now well understood that dynamical processing in dense star clusters similar in properties to the 
GCs can be a dominant formation channel for the BBHs observed by LIGO \citep{Banerjee2010,Ziosi2014,Rodriguez2015,Rodriguez2016a,Chatterjee2017a,Chatterjee2017b}. It has also been shown that the retention fraction of BHs can lead to drastic 
differences in the way the host cluster evolves \citep{Chatterjee2017a}, which in turn 
affects the overall BH dynamics and BBH formation. 

Recent simulations find that the binary fraction in retained BHs (with 
BH or non-BH companions) remains low \citep{Morscher2013,Morscher2015,Chatterjee2017a}. In addition, 
the duty cycle for the active state of a BH-X-ray binary (BH-XRB) is low \citep{Kalogera2004}. Thus, 
it has been argued that finding even a few BH-XRB candidates in GCs likely indicates 
much larger populations of retained BHs in those GCs in undetectable configurations \citep[e.g.,][]{Umbreit2012}. 

In this work, we explore the formation of mass-transferring BH binaries (MTBHBs) in GCs and examine what the presence of a MTBHB implies about the total population of BHs within a cluster. In Section \ref{sec:method}, we describe our computational method, numerical 
setup, and initial conditions. We also explain how we identify an accreting BH in a snapshot of our models, which we 
use as a proxy to BHs detectable in that cluster via, e.g., X-ray or radio 
observations. In Section \ref{sec:results}, we discuss our key results and examine the properties of the MTBHBs found in our models. In Section \ref{sec:rates} we explain some of the key results using rate-based analysis. While all of the previous sections focus on understanding the 
formation and properties of the MTBHBs that are 
retained in GCs today, we devote Section\ \ref{sec:ejected} to briefly discuss the key properties of MTBHBs 
that have been ejected from the cluster. We conclude in Section \ref{sec:conclusion}.

\section{Method}
\label{sec:method}
We model massive star clusters using our H\'{e}non-style Monte Carlo cluster dynamics code, \cmc, developed and 
extensively tested by Northwestern's cluster dynamics group 
\citep[e.g.,][]{Joshi2000,Joshi2001,Fregeau2003, Fregeau2007, Chatterjee2010,Chatterjee2013,Umbreit2012,Morscher2013,Rodriguez2016b}. The most recent detailed description of \cmc\ 
and its validation can be found in \citet{Pattabiraman2013,Morscher2015,Rodriguez2016b}.

In addition  to two-body relaxation (the primary driver of 
evolution in high-$N$ collisional gravitational systems) \cmc\ incorporates all the relevant physical processes for studying the formation and evolution of binary systems containing BHs. We model binary-mediated gravitational scattering encounters explicitly using the {\tt Fewbody} small-$N$ integrator \citep{Fregeau2004}. 
Single and binary stellar evolution are implemented using the \sse\ and \bse\ 
software packages 
\citep{Hurley2000,Hurley2002,Chatterjee2010,Kiel2009} modified to incorporate our latest understanding of the BH mass function, stellar winds, and natal kicks due to supernovae \citep[SN; e.g.,][]{Vink2001,Fryer2001,Belczynski2002}. Physical collisions are included and the properties 
of the collision products are also obtained using \sse\ prescriptions.

\begin{deluxetable*}{ccccccc|cc|ccc||cc}
\tabletypesize{\scriptsize}
\tablewidth{0pt}
\tablecaption{List of model properties \label{table:params}}
\tablehead{
	\colhead{No.} &
    \colhead{$N$} &
    \colhead{$r_{\rm{G}}$} &
    \colhead{$w_o$} &
    \colhead{$f_b$} &
    \colhead{$Z$} &
    \colhead{$r_v$} &
    \multicolumn{2}{c}{BH-formation kick} &    
    \multicolumn{3}{c}{High Mass Binaries} &
    \colhead{$N_{\rm{BH}}^{\rm{tot}}$} &
    \colhead{$N_{\rm{MTBHB}}$} \\
    \cline{8-12}\\
    \colhead{} &
    \colhead{$(10^5$)} &
    \colhead{$(\rm{kpc})$} &
    \colhead{} &
    \colhead{} &
    \colhead{} &
    \colhead{$(\rm{pc})$} &
    \colhead{$\frac{\sigma_{\rm{BH}}}{\sigma_{\rm{NS}}}$} &
    \colhead{\texttt{FB}} &
    \colhead{$f_{b,\rm{high}}$} &
    \colhead{$q$ range} &
    \colhead{$\frac{dn}{d\, \log P}$} &
    \colhead{} &
    \colhead{}
}
\startdata
1 & 1 & 4.6 & 5 & 0.05 & 0.00055 & 1 & 1 & y & 0.05 & $[0.1/m_p,\,1]$ & $P^0$ & 0-1 & 0\\
2 & 2.4 & 4.6 & 5 & 0.05 & 0.00055 & 0.8 & 1 & y & 0.05 & $[0.1/m_p,\,1]$ & $P^0$  & 5 & 0 \\
3 & 2.4 & 4.6 & 5 & 0.05 & 0.00055 & 1 & 1 & y & 0.05 & $[0.1/m_p,\,1]$ & $P^0$ & 2-10 & 0-1 \\
4 & 2.6 & 4.6 & 5 & 0.05 & 0.00055 & 0.8 & 1 & y & 0.05 & $[0.1/m_p,\,1]$ & $P^0$ & 0-8 & 0-1 \\
5 & 2.6 & 4.6 & 5 & 0.05 & 0.00055 & 1.2 & 1 & y & 0.05 & $[0.1/m_p,\,1]$ & $P^0$ & 1-13 & 0-1 \\
6 & 2.6 & 4.6 & 5 & 0.05 & 0.00055 & 1 & 1 & y & 0.05 & $[0.1/m_p,\,1]$ & $P^0$ & 2-11 & 0 \\
7 & 3.25 & 4.6 & 5 & 0.05 & 0.00055 & 1 & 1 & y & 0.05 & $[0.1/m_p,\,1]$ & $P^0$ & 3-19 & 0-1 \\
8 & 3.5 & 4.6 & 4 & 0.05 & 0.00055 & 1 & 1 & y & 0.05 & $[0.1/m_p,\,1]$ & $P^0$ & 2-17 & 0 \\
9 & 3.5 & 4.6 & 5 & 0.05 & 0.00055 & 1 & 1 & y & 0.05 & $[0.1/m_p,\,1]$ & $P^0$ & 5-19 & 0 \\
10 & 3.75 & 4.6 & 4 & 0.05 & 0.00055 & 1 & 1 & y & 0.05 & $[0.1/m_p,\,1]$ & $P^0$ & 3-21 & 0-1 \\
11 & 3.75 & 4.6 & 5.5 & 0.05 & 0.00055 & 0.8 & 1 & y & 0.05 & $[0.1/m_p,\,1]$ & $P^0$ & 0-14 & 0-1 \\
12 & 3.75 & 4.6 & 5 & 0.05 & 0.00055 & 0.8 & 1 & y & 0.05 & $[0.1/m_p,\,1]$ & $P^0$ & 0-17 & 0-1 \\
13 & 3.75 & 4.6 & 5 & 0.05 & 0.00055 & 1 & 1 & y & 0.05 & $[0.1/m_p,\,1]$ & $P^0$ & 3-21 & 1 \\
14 & 3.75 & 4.6 & 6 & 0.05 & 0.00055 & 1 & 1 & y & 0.05 & $[0.1/m_p,\,1]$ & $P^0$ & 9-29 & 0-1 \\
15 & 3.8 & 4.6 & 5.1 & 0.05 & 0.00055 & 0.9 & 1 & y & 0.05 & $[0.1/m_p,\,1]$ & $P^0$ & 1-17 & 0-2 \\
16 & 3.8 & 4.6 & 5.2 & 0.05 & 0.00055 & 0.85 & 1 & y & 0.05 & $[0.1/m_p,\,1]$ & $P^0$ & 1-18 & 0 \\
17 & 3 & 4.6 & 4 & 0.05 & 0.00055 & 2 & 1 & y & 0.05 & $[0.1/m_p,\,1]$ & $P^0$ & 10-53 & 0 \\
18 & 3 & 4.6 & 5 & 0.05 & 0.00055 & 1 & 1 & y & 0.05 & $[0.1/m_p,\,1]$ & $P^0$  & 3-12 & 0-1 \\
19 & 4 & 4.6 & 5 & 0.05 & 0.00055 & 0.8 & 1 & y & 0.05 & $[0.1/m_p,\,1]$ & $P^0$ & 2-17 & 1-2 \\
20 & 4 & 4.6 & 5 & 0.05 & 0.00055 & 1 & 1 & y & 0.05 & $[0.1/m_p,\,1]$ & $P^0$ & 5-26 & 0 \\
21 & 4 & 4.6 & 6 & 0.05 & 0.00055 & 1 & 1 & y & 0.05 & $[0.1/m_p,\,1]$ & $P^0$ & 5-24 & 0-1 \\
22 & 6 & 4.6 & 4 & 0.05 & 0.00055 & 1 & 1 & y & 0.05 & $[0.1/m_p,\,1]$ & $P^0$ & 17-68 & 0-2 \\
23 & 6 & 4.6 & 5 & 0.05 & 0.00055 & 1 & 1 & y & 0.05 & $[0.1/m_p,\,1]$ & $P^0$  & 16-57 & 0-1 \\
24 & 6 & 4.6 & 6 & 0.05 & 0.00055 & 1 & 1 & y & 0.05 & $[0.1/m_p,\,1]$ & $P^0$  & 6-27 & 1-2 \\
25 & 8 & 4.6 & 5 & 0.05 & 0.00055 & 1 & 1 & y & 0.05 & $[0.1/m_p,\,1]$ & $P^0$  & 37-140 & 0 \\
26 & 8 & 8 & 5 & 0.05 & 0.001 & 2 & 1 & y & 0.05 & $[0.1/m_p,\,1]$ & $P^0$  & 464-643 & 0 \\
27 & 8 & 8 & 5 & 0.05 & 0.001 & 2 & 1 & n & 0.05 & $[0.1/m_p,\,1]$ & $P^0$  & 4 & 0 \\
28 & 8 & 8 & 5 & 0.05 & 0.001 & 2 & 0.1 & n & 0.05 & $[0.1/m_p,\,1]$ & $P^0$ & 241-344 & 1 \\
29 & 8 & 8 & 5 & 0.05 & 0.001 & 2 & 0.01 & n & 0.05 & $[0.1/m_p,\,1]$ & $P^0$ & 739-957 & 0 \\
30 & 8 & 8 & 5 & 0.04 & 0.001 & 2 & 1 & y & 0.7 & $[0.6,1]$ & $P^{-0.55}$ & 61-111 & 0-1 \\
31 & 8 & 8 & 5 & 0.04 & 0.001 & 2 & 1 & n & 0.7 & $[0.6,1]$ & $P^{-0.55}$ & 0 & 0 \\
32 & 8 & 8 & 5 & 0.04 & 0.001 & 2 & 0.1 & n & 0.7 & $[0.6,1]$ & $P^{-0.55}$ & 39-60 & 0-1 \\
33 & 8 & 8 & 5 & 0.04 & 0.001 & 2 & 0.01 & n & 0.7 & $[0.6,1]$ & $P^{-0.55}$ & 126-190 & 0 \\
34 & 8 & 8 & 5 & 0.05 & 0.001 & 2 & 1 & y & 0.05 & $[0.1/m_p,\,1]$ & $P^0$  & 464-665 & 1-2 \\
35 & 8 & 8 & 5 & 0.05 & 0.001 & 2 & 1 & n & 0.05 & $[0.1/m_p,\,1]$ & $P^0$ & 6-8 & 0-1 \\
36 & 8 & 8 & 5 & 0.05 & 0.001 & 2 & 0.01 & n & 0.05 & $[0.1/m_p,\,1]$ & $P^0$ & 759-949 & 0 \\
37 & 8 & 8 & 5 & 0.05 & 0.001 & 2 & 1 & y & 1 & $[0.1/m_p,\,1]$ & $P^0$ & 467-625 & 0 \\
38 & 8 & 8 & 5 & 0.05 & 0.001 & 2 & 1 & y & 0 & $-$ & $-$ & 458-642 & 2-3 \\
39 & 8 & 8 & 5 & 0.05 & 0.001 & 2 & 1 & y & 0.7 & $[0.1/m_p,\,1]$ & $P^{-0.55}$  & 437-616 & 0 \\
40 & 8 & 8 & 5 & 0.05 & 0.001 & 2 & 1 & n & 0.7 & $[0.1/m_p,\,1]$ & $P^{-0.55}$  & 18-34 & 1-2 \\
41 & 8 & 8 & 5 & 0.05 & 0.001 & 2 & 0.1 & n & 0.7 & $[0.1/m_p,\,1]$ & $P^{-0.55}$  & 279-393 & 0 \\
42 & 8 & 8 & 5 & 0.05 & 0.001 & 2 & 0.01 & n & 0.7 & $[0.1/m_p,\,1]$ & $P^{-0.55}$  & 673-853 & 0 \\
43 & 8 & 8 & 5 & 0.05 & 0.001 & 2 & 1 & y & 0.7 & $[0.6,1]$ & $P^{-0.55}$  & 503-702 & 0 \\
44 & 8 & 8 & 5 & 0.05 & 0.001 & 2 & 1 & n & 0.7 & $[0.6,1]$ & $P^{-0.55}$  & 31-48 & 0-1 \\
45 & 8 & 8 & 5 & 0.05 & 0.001 & 1 & 0.1 & n & 0.7 & $[0.6,1]$ & $P^{-0.55}$  & 338-444 & 1 \\
46 & 8 & 4 & 5 & 0.05 & 0.001 & 2 & 1 & y & 0.7 & $[0.6,1]$ & $P^{-0.55}$  & 462-656 & 1 \\
47 & 8 & 4 & 5 & 0.05 & 0.001 & 2 & 1 & n & 0.7 & $[0.6,1]$ & $P^{-0.55}$ & 31-46 & 1-3 \\
48 & 8 & 4 & 5 & 0.05 & 0.001 & 2 & 0.1 & n & 0.7 & $[0.6,1]$ & $P^{-0.55}$ & 315-434 & 0-1 \\
49 & 8 & 4 & 5 & 0.05 & 0.001 & 2 & 0.01 & n & 0.7 & $[0.6,1]$ & $P^{-0.55}$ & 647-863 & 0-2 \\
50 & 8 & 2 & 5 & 0.05 & 0.001 & 2 & 1 & n & 0.7 & $[0.6,1]$ & $P^{-0.55}$ & 13-47 & 0 \\
\enddata
\tablecomments{Relevant properties of all GC models used in this study. Each initial parameter is described in Section\ \ref{S:model_properties}. FB denotes whether BH natal kicks are fallback dependent or not. 
Ranges in $N_{\rm{MTBHB}}$ and $N_{\rm{BH}}^{\rm{tot}}$ are from all model snapshots 
with $t\geq 8\,\gyr$. 
}
\end{deluxetable*}

\subsection{Model properties}\label{S:model_properties}
We use $50$ different GC models of varying initial structural properties, listed in Table \ref{table:params}. We vary the initial number of objects in the cluster ($N$), the initial galactocentric distance ($r_{\rm{G}}$), the King concentration parameter ($w_o$), the overall primordial binary fraction ($f_b$), the initial virial radius ($r_v$), and the initial metallicity ($Z$). The stellar masses (primary mass for a primordial binary) are sampled from the IMF presented in \citet{Kroupa2001} in the range $0.1$--$100\,\msun$. An appropriate number of stars are then randomly chosen based on the adopted $f_b$ and $N$ for the model. Secondaries are assigned to these stars based on a flat distribution in mass ratios ($q\equiv m_s/m_p$, where $m_s$ and $m_p$ denote the secondary and primary masses, respectively). The initial orbital periods ($P$) are drawn from a distribution of the form $dn/d\log P\propto P^\alpha$. The initial eccentricities are thermal.  

All core-collapsed neutron stars get birth kicks drawn from a Maxwellian distribution with $\sigma=\sigma_{\rm{NS}}=265\, \rm{km\,s}^{-1}$ \citep{Hobbs2005}. We use four separate prescriptions to obtain BH kick magnitudes. In the first prescription, we assume BHs are formed with significant fallback and calculate the natal kicks by sampling from the same kick distribution as the neutron stars, but reduced in magnitude according the the fractional mass of the fallback material 
\citep[see][for more details]{Morscher2015}. In the other three variations we neglect fallback and simply use $\sigma_{\rm{BH}}=\sigma_{\rm{NS}}$, $\sigma_{\rm{BH}}=0.1\,\sigma_{\rm{NS}}$, and $\sigma_{\rm{BH}}=0.01 \, \sigma_{\rm{NS}}$. 

In some models we specifically vary the initial binary fraction, $f_{b,\rm{high}}$, for high-mass ($>15\,\msun$) stars independent of the overall binary fraction. In these models we also vary the range in $q$ and the initial period distribution for the high-mass stars motivated by the 
observational constraints from \citet[][]{Sana2012}. 


In each model we simultaneously track the dynamical evolution and the single and binary stellar evolution. 
We then use these models to extract the number, properties, and dynamical history of MTBHBs in any snapshot 
and compare them with the number of retained BHs in the cluster at that time. Since we are interested in MTBHBs in GCs, we focus on snapshots older than $\sim8\,\gyr$. 


\subsection{Identifying Mass-Transferring BH Binaries}
\label{sec:rhoche-lobe}
To calculate the Roche-lobe radius of the components in the binary system, \bse\ adopts the formula introduced in \citet{Eggleton1983}:

\begin{equation}
\label{eq:RL}
R_L = a\frac{0.49q^{2/3}}{0.6q^{2/3}+log(1+q^{1/3})}
\end{equation}
where $a$ is the semi-major axis of the binary and 
$q\equiv M_D/M_A$ is the mass ratio of the donor to the accretor.

It should be noted that in a GC, where eccentricity-introducing dynamical interactions are common, it is likely that a substantial fraction of mass transferring binaries will begin mass transfer during periastron passage while significant 
eccentricity is still present. For such systems, a modified calculation of the Roche lobe which includes binary eccentricity may be more appropriate. 

In order to attain a sense of the importance of eccentricity in the determination of mass-transferring systems in our models, we also utilized a modified version of Equation (\ref{eq:RL}) which uses the periastron distance, $a(1-e)$, in place of $a$. In this case, the total number of MTBHBs found in our models increased by only $\sim 8 \%$ relative to the number obtained using the default prescription in \bse. This increase is mainly due to the fact 
that if eccentricity is considered, potentially mass-transferring systems start transferring mass slightly earlier than if eccentricity is ignored. We also find that this modest increase is not correlated with the retained number of BHs and other structural properties of the cluster. 
Therefore, to remain consistent with the treatments of 
mass-transferring systems in our models using \cmc, we adopt the definition in Equation (\ref{eq:RL}), and note that implementing eccentricity into the Roche-lobe calculation is likely to only lead to a modest increase in the total number of MTBHBs at any given time in the cluster models.

\section{Results} \label{sec:results}

\begin{figure}[t!]
\plotone{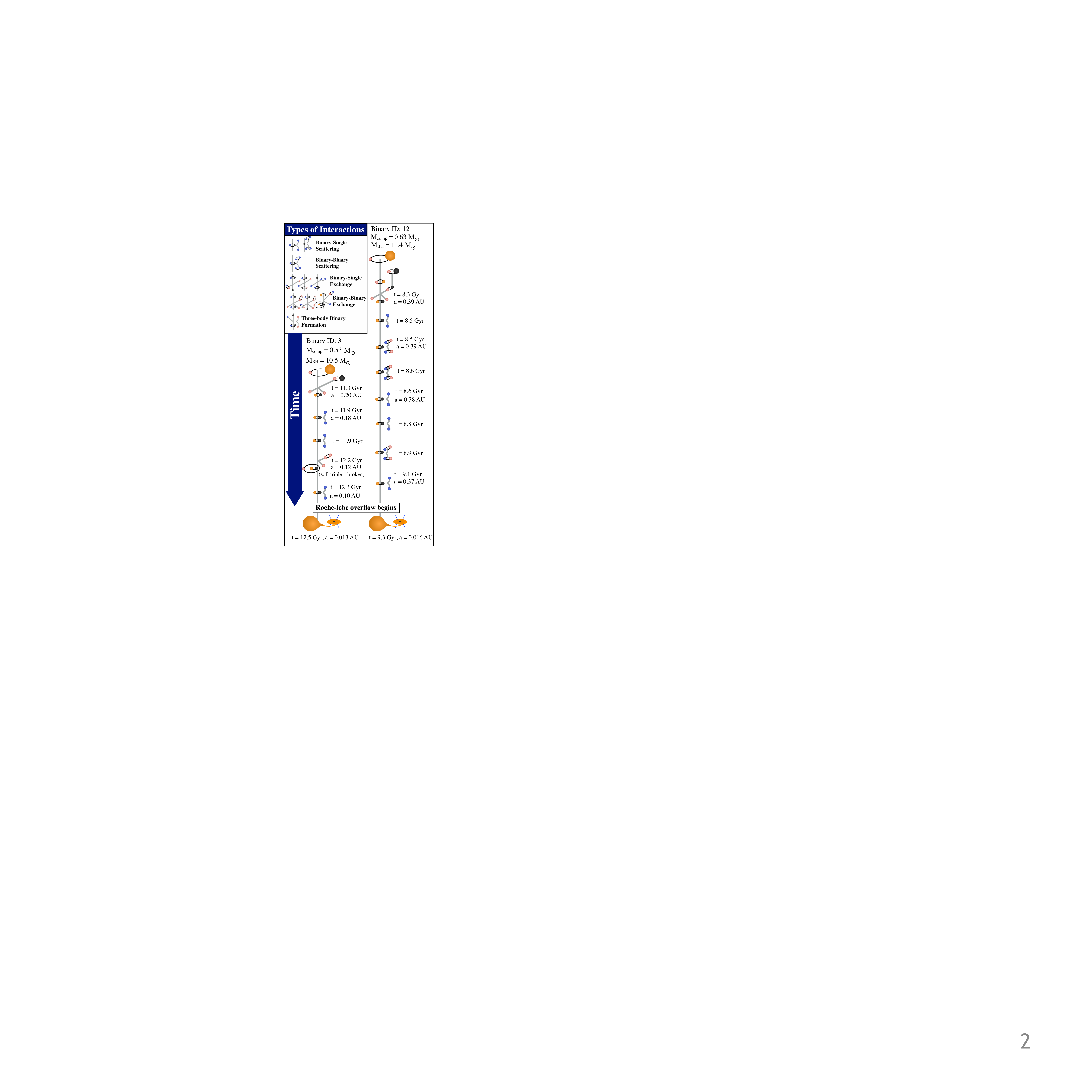}
\caption{\label{fig:formation} \footnotesize Illustration of the dynamical interactions experienced by two example binaries from the formation of each binary to the onset of Roche-lobe overflow. Symbols for different types of dynamical interactions are illustrated in the inset. }
\end{figure}

Using the definition of Roche-lobe overflow described in Equation\ (\ref{eq:RL}), we search our 50 GC models for MTBHBs. As we are interested only in MTBHBs  that may be observable at the present day, we limit our search to late stages of cluster evolution, defined here as snapshots in time with $t \geq$ 8 Gyr. Note that the choice of 8 Gyr is arbitrary and meant to reflect 
the approximate lower limit of GC ages in the MW.

Each model contains $\sim100-1000$ snapshots in time spaced $\sim10-100$ Myr apart. Each of these snapshots serves as a unique representation of a GC at a different point in its evolution.

In our 50 models, we find a total of 17 MTBHBs, whose orbital parameters are shown in Table \ref{table:2}. For these 17 systems, 14 of the companion stars are main sequence stars (MS), 2 are giants (G), and 1 is a white dwarf (WD). (Note that recent constraints on the companion of the BH candidate X9 in 47 Tuc \citep{Bahramian2017} suggest a Carbon-Oxygen WD donor.) The largest number of total MTBHBs found in any model at a single snapshot in time is 2. Three independent cluster models (model numbers 24, 38, and 49 in Table \ref{table:params}) contain 2 MTBHBs for $t \geq 8\,\rm{Gyr}$. Two of these three models have initial $N=8 \times 10^5$, and the third has initial $N=6 \times 10^5$. These three models contain 
$\sim 650$ (model 49), $\sim 550$ (model 38), and $\sim 20$ (model 24) retained BHs, respectively, at the time the MTBHBs are found.

Each of these 17 systems is formed as a result of dynamical encounters. Although our models initially have $f_b=4$--$5\%$, and up to $100\%$ primordial binary fraction for high mass stars, none of the MTBHBs in these models at $t\geq8\, \gyr$ are found to be primordial. This is because the BH binaries in our models repeatedly change companions. Each BH in the 17 identified MTBHBs at late times had many interactions and exchanges before acquiring the final companion which ultimately fills its Roche Lobe.

Dynamical encounters also play a critical role in the evolution of these binary systems once they are formed. Interactions alter the semi-major axis and eccentricity of the binaries involved, which ultimately determine whether or not the systems become Roche-lobe overflowing.


\begin{deluxetable*}{c|c|c|cc|c|ccc}
\tabletypesize{\scriptsize}
\tablecolumns{9}
\tablewidth{0pt}
\tablecaption{MTBHB orbital parameters \label{table:2}}
\tablehead{
	\colhead{Binary ID} &
    \colhead{$t_{\rm{MT}}$} &
    \colhead{Comp. Type} &
    \colhead{$M_{\rm{BH}}$} &
    \colhead{$M_{\rm{comp}}$} &
    \colhead{$a$} &
    \colhead{$\frac{\Delta a_{\rm{bse}}}{\Delta a_{\rm{tot}}}$} &
    \colhead{$\frac{\Delta a_{\rm{BB}}}{\Delta a_{\rm{tot}}}$} &
    \colhead{$\frac{\Delta a_{\rm{BS}}}{\Delta a_{\rm{tot}}}$} \\
    \cline{4-5}
    \colhead{} & 
    \colhead{($\gyr$)} &
    \colhead{} &
    \multicolumn{2}{c}{($\msun$)} &
    \colhead{(AU)} &
    \colhead{} &
    \colhead{} &
    \colhead{}
}
\startdata
1 & 11.1 & G & 10.2 & 1.02 & 0.61 & 0 & 0 & 1.0 \\
2 & 7.73 & MS & 13.8 & 0.39 & 0.01 & 0.98 & $2.23 \times 10^{-2}$ & 0  \\
3 & 12.5 & MS & 10.5 & 0.53 & 0.013 & 0.52 & 0.30 & 0.18 \\
4 & 2.75 & MS & 10.2 & 0.17 & 0.007 & 0.98 & $1.72 \times 10^{-2}$ & 0 \\
5 & 0.47 & MS & 11.6 & 0.23 & 0.009 & 1.0 & 0 & 0 \\
6 & 14.1 & MS & 7.92 & 0.48 & 0.011 & 0 & 0 & 0 \\
7 & 8.12 & G & 6.20 & 0.93 & 0.124 & 1.0 & 0 & 0  \\
8 & 2.64 & MS & 10.8 & 0.52 & 0.013 & 0.79 & 0.21 & 0 \\
9 & 8.41 & MS & 11.9 & 0.74 & 0.020 & 0 & 0 & 0  \\
10 & 2.95 & MS & 11.0 & 0.39 & 0.011 & 1.0 & 0 & 0 \\
11 & 1.61 & MS & 7.44 & 0.15 & 0.006 & 1.0 & 0 & 0 \\
12 & 9.35 & MS & 11.4 & 0.63 & 0.016 & 0.95 & $2.05 \times 10^{-3}$ & $4.31 \times 10^{-2}$ \\
13 & 5.34 & MS & 3.19 & 0.34 & 0.007 & 1.0 & 0 & 0 \\
14 & 0.24 & WD & 3.28 & 0.02 & 0.003 & 0 & 0 & 0 \\
15 & 0.66 & MS & 9.56 & 0.28 & 0.009 & 1.0 & 0 & 0 \\
16 & 9.47 & MS & 11.9 & 0.90 & 0.020 & 0 & 0 & 0 \\
17 & 3.53 & MS & 11.7 & 0.11 & 0.006 & 1.0 & $2.05 \times 10^{-4}$ & 0 \\
\enddata
\tablecomments{Orbital parameters for each MTBHB identified in our GC models at the time of the onset of mass-transfer.}
\end{deluxetable*}

\subsection{Types of dynamical interactions}
\label{interactions}

The relevant dynamical interactions experienced by our 17 MTBHBs can be split into two classes: interactions between a binary and a single star (binary--single) and interactions between a binary with another binary (binary--binary). Once the star identified as the donor star in a MTBHB is acquired by the identified BH, the binary--single and binary--binary interactions considered are all, by definition, non-exchange encounters. Before the MTBHB is formed, however, exchange encounters are common and each BH accretor goes through many exchange encounters before it acquires the donor which eventually fills its Roche lobe.

Figure \ref{fig:formation} illustrates the dynamical evolution of two example MTBHBs (Binary 3 and Binary 12 from Table \ref{table:2}) from the formation of the binary to the onset of Roche-lobe overflow. In particular, this figure illustrates the various types of dynamical interactions the binaries experience and the effect of these interactions on the separation of the binary.

One possible outcome of a binary--binary interaction is the formation of a hierarchical triple system where a tight binary exchanges into a wide binary, ejecting the single companion of the latter binary. 
Since we cannot treat triple stellar evolution consistently and to limit computational cost, we break such triples within \cmc. Here, we identify 
MTBHBs where the orbits were never significantly altered ($\Delta a/a\lesssim5\%$) due to the formation (and artificial breaking) of a transient triple and explore their properties and formation history first. Effects of triples on the formation of MTBHBs are discussed separately 
in Section\ \ref{triples}.

In addition to dynamical interactions, the binaries also evolve due to effects of standard binary star evolution. Of particular relevance here are tidal interactions, which circularize and shrink the orbit over time. We utilize the tidal treatment implemented in \texttt{BSE}, which in turn uses the treatment of \citet{Hut1981} to calculate the effect of tides on the evolution of the semi-major axis and eccentricity of binaries. 

\subsection{Dynamical history of mass-transferring BH binaries}
\label{sec:history}
\begin{figure}
\plotone{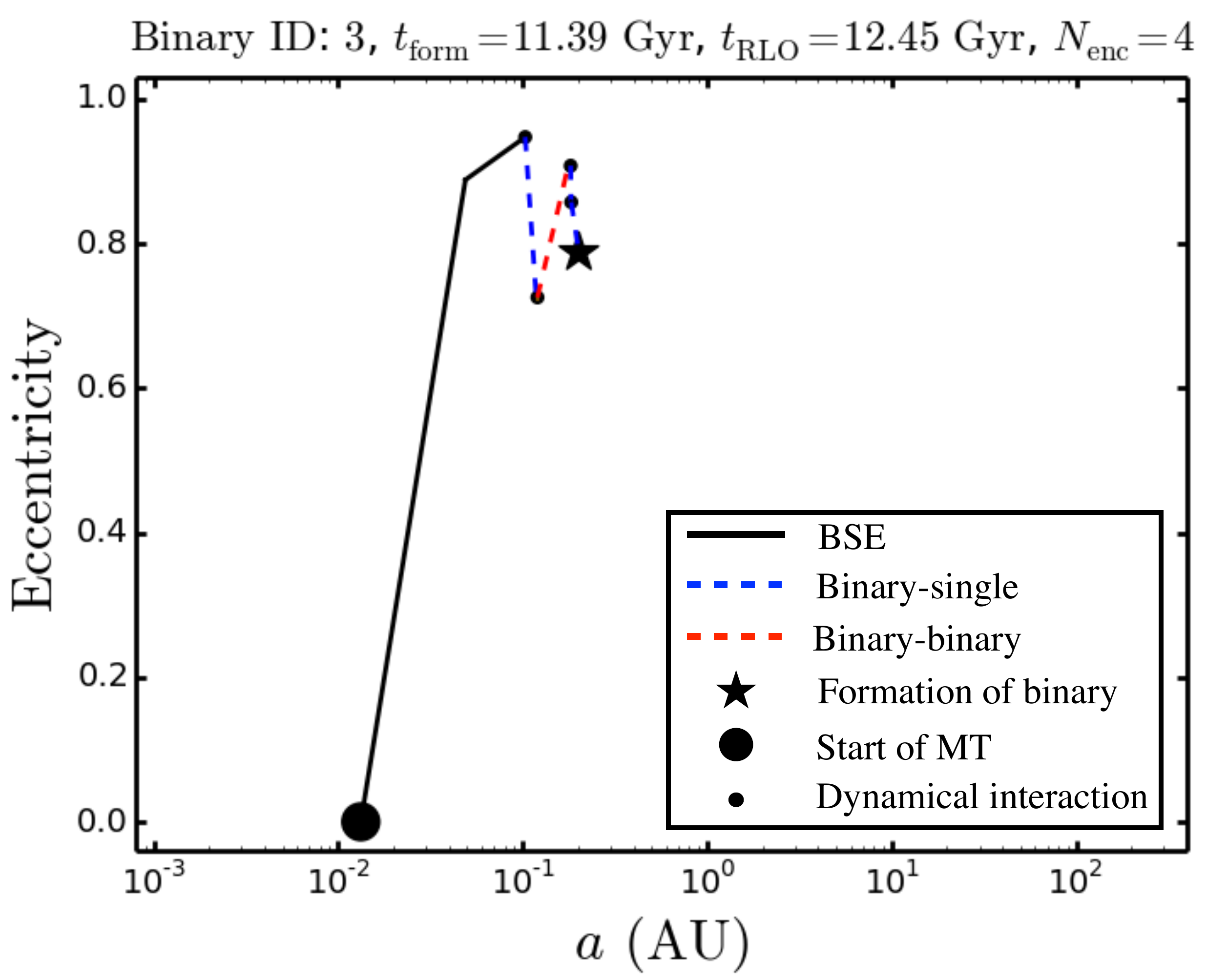}
\caption{\label{fig:7} \footnotesize Evolution of Binary 3 (Table \ref{table:2}) in the $a$ vs $e$ plane from the time of formation of the binary (marked by the black star) to the onset of Roche-lobe overflow (marked by the large black dot). Each small dot represents individual dynamical interactions. The solid black line shows evolution due to standard binary star evolution. Dashed lines show evolution from dynamical encounters. Dashed red indicates binary--binary interactions and dashed blue denotes binary--single interactions.}
\end{figure}
\begin{figure}
\plotone{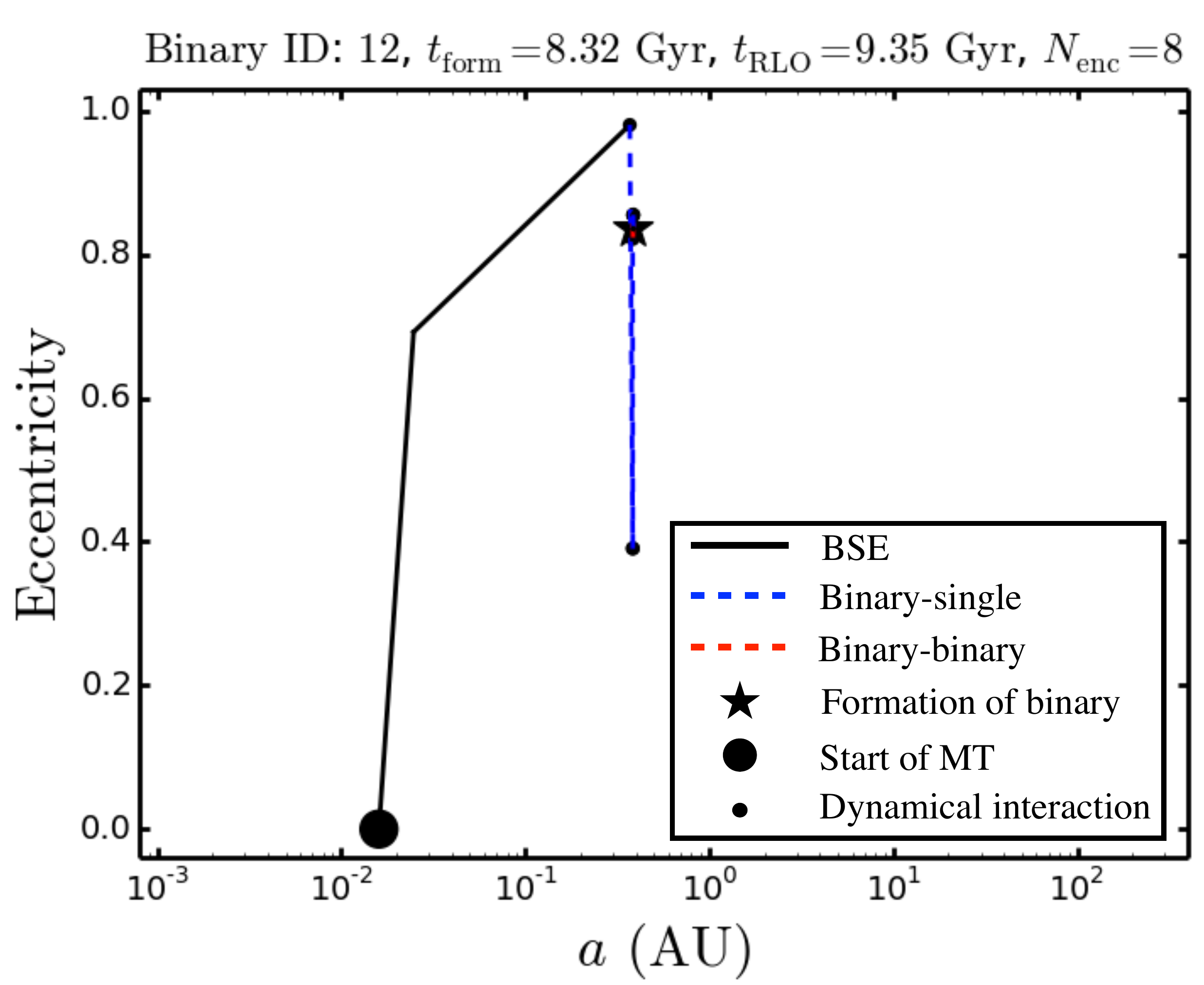}
\caption{\label{fig:21} \footnotesize Same as Figure \ref{fig:7} but for the evolution of Binary 12 (Table \ref{table:2}).}
\end{figure}

Figures \ref{fig:7} and \ref{fig:21} illustrate the evolution of Binary 3 and Binary 12 (see Table \ref{table:2}) in the $a$ vs $e$ plane from the time the binary is formed (marked by the solid black star) to the onset of Roche-lobe overflow (RLO; marked by the large black dot). In both figures, the small dots represent single dynamical encounters. The solid black line shows evolution due to standard binary star evolution (tidal interactions, etc.) as calculated using \bse. 
Dashed red indicates binary--binary interactions and dashed blue indicates binary--single interactions. Note that similar figures for all of the model binaries are available in the online version of the paper.

Note that the abrupt changes in slope for the orbital decay curves in Figures \ref{fig:7} and \ref{fig:21} for the binary star evolution (solid black lines) are an artifact of the discrete snapshots in time taken by \cmc. Actual changes in slope resulting from standard binary star evolution mechanisms such as tidal decay are expected to be continuous \citep[e.g.,][]{Hut1981} in contrast to changes resulting from dynamical interactions, which will be discontinuous.

Each of the 17 identified MTBHBs are driven to mass-transfer through one of three distinct scenarios. In the first scenario, a dynamical interaction (or series of interactions) induces a high orbital eccentricity. This high eccentricity initiates tidal decay (illustrated by the solid black lines of Figures \ref{fig:7} and \ref{fig:21}), which ultimately drives the binary to the point of mass-transfer. This scenario is illustrated by the two binaries shown in Figures \ref{fig:7} and \ref{fig:21} as well as binaries 5, 7, 8, 10, 13, and 15 from Table \ref{table:2}. Note that binaries 5, 7, and 10 are dynamically assembled with a high eccentricity and tidal decay alone hardens the binary to the point of mass-transfer once the binary has formed.

In the second scenario, the MTBHB is assembled through a dynamical exchange in a much tighter ($\sim 10^{-2}$ AU) configuration and undergoes a combination of tidal decay and dynamical hardening that ultimately induces mass-transfer. This scenario describes binaries 2, 4, 11, and 17.

For some of these (binaries 5, 7, 10, 11, 12, 15, and 17), the dynamical hardening has only small or negligible effect, as seen in columns 8 and 9 of Table \ref{table:2}. However, the role of dynamics cannot be underestimated for these binaries, because it was a dynamical encounter that assembled these binaries in the first place.

In the third scenario, the binary is assembled as a Roche-lobe overflowing system. Binaries 6, 9, 14, and 16 of Table \ref{table:2} fall into this category. Because they begin mass-transfer immediately upon formation, the $\Delta a_i$ values shown in columns 7-9 of Table \ref{table:2} are all zero for these four binaries.

\begin{figure}
\plotone{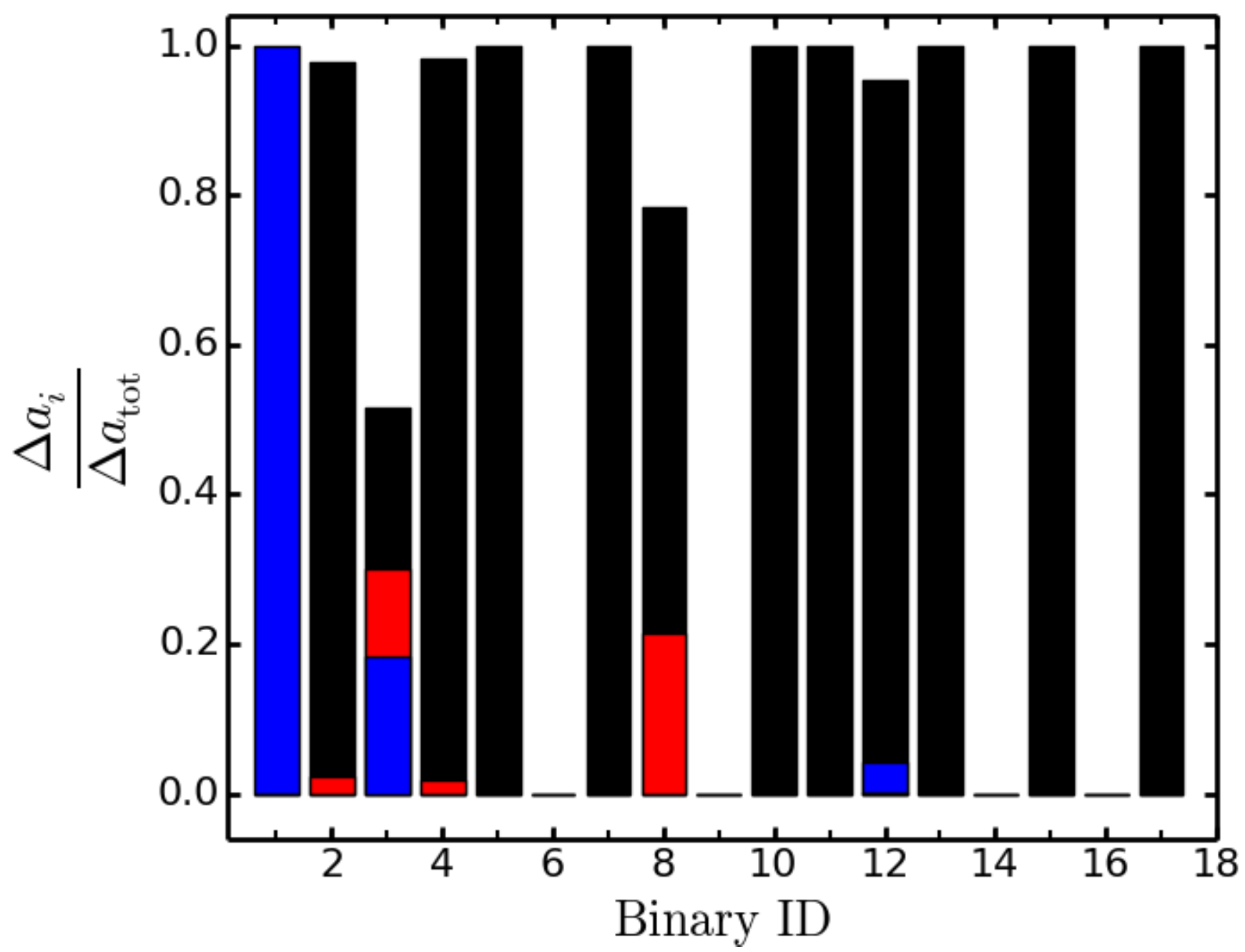}
\caption{\label{fig:deltaA} \footnotesize {Relative importance of different evolutionary components on the overall evolution of the semi-major axis for each of our 17 model binaries. For each binary, black shows the change in $a$ due to binary star evolution ($\Delta a_{\rm{bse}}/\Delta a_{\rm{tot}}$), red shows the change due to binary--binary interactions in which a triple is not formed ($\Delta a_{\rm{BB}}/\Delta a_{\rm{tot}}$), and blue shows the change due to binary--single interactions ($\Delta a_{\rm{BS}}/\Delta a_{\rm{tot}}$). Because binaries 6, 9, 14, and 16 are assembled as Roche-lobe overflowing systems, $\Delta a_i = 0$ for all processes.}}
\end{figure}

Figure \ref{fig:deltaA} shows the relative importance of each evolutionary component toward the overall change in the semi-major axis for each of the 17 model binaries before RLO begins. For each binary, the change in $a$ due to each process: binary star evolution (black), binary--binary interactions (red), and binary--single interactions (blue), is scaled against the total change in $a$ since formation, $\Delta a_{\rm{tot}}$.  As illustrated by Figure \ref{fig:deltaA}, binary star evolution, in particular, tidal decay of the semi-major axis, dominates the evolution of $a$ for most binaries while dynamical encounters are important to form the binary and then harden it enough 
for tidal decay to bring the binary to RLO. 

\begin{figure}
\plotone{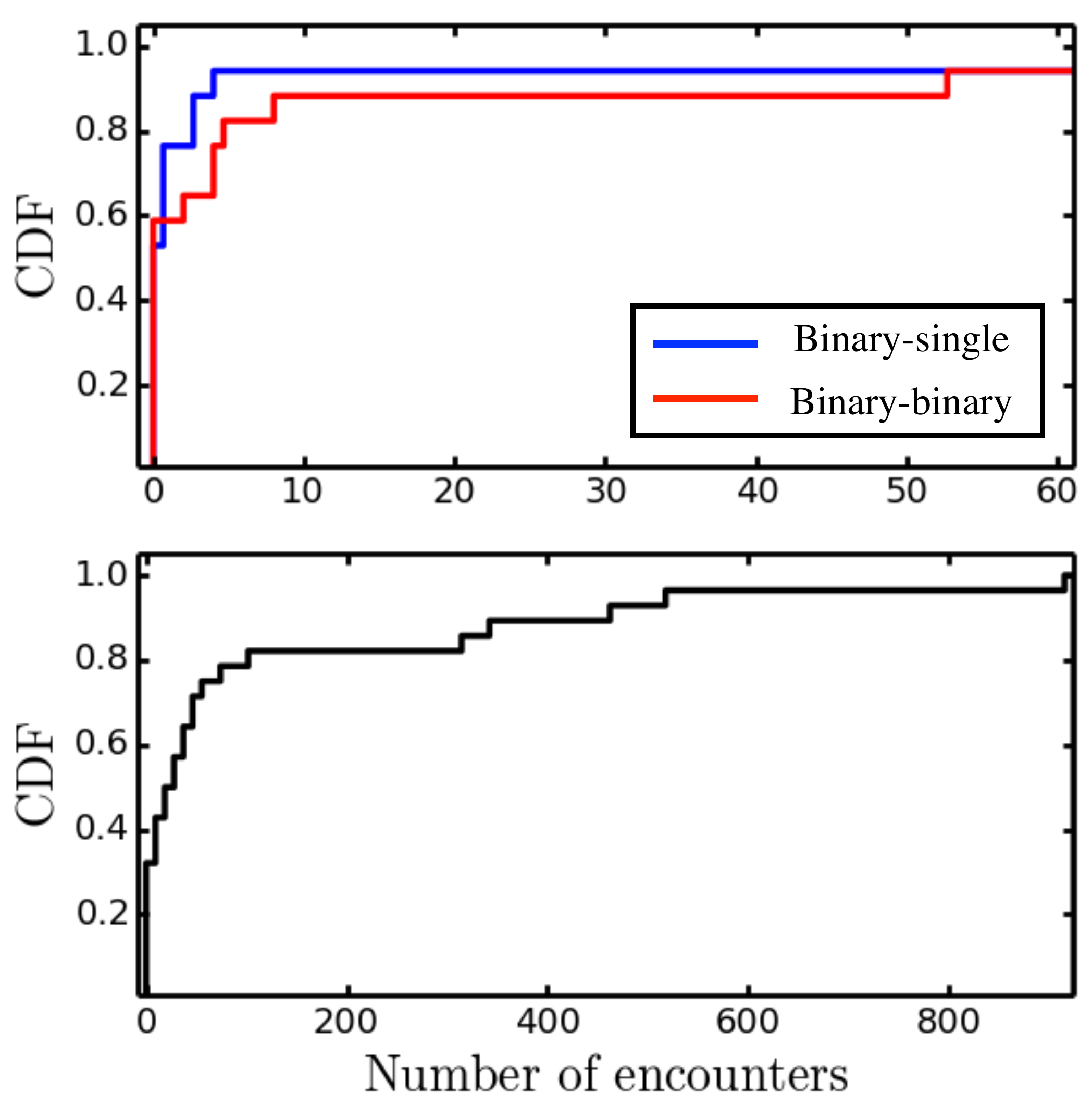}
\caption{\label{fig:CDF} \footnotesize Top panel shows the cumulative distribution of the total number of encounters each binary experiences from the time the binary is formed to the time Roche-lobe overflow begins. Red shows the number of binary--binary encounters and blue shows the number of binary--single encounters. Bottom panel shows the cumulative distribution of the total number of encounters experienced by each BH component in mass-transferring binaries throughout its complete history (before and after the final MTBHB is formed) up until the onset of Roche-lobe overflow.}
\end{figure}

Figure \ref{fig:CDF} shows the number of dynamical encounters the 17 MTBHBs experience throughout their evolution. The top panel shows the cumulative distribution function (CDF) of the number of encounters of each type that each system experiences from the time the binary is formed to the time RLO begins. The blue line shows the number of binary--single encounters and the red line shows the number of binary--binary encounters. The bottom panel shows the total number of encounters of all types experienced by the BHs in these systems. While the median value for the total number of encounters 
is 1 after the formation of the MTBHB, the median number of total encounters the 
BHs experience (before and after the formation of the mass-transferring system) is 29. 

Once a sufficiently hard binary with a BH and a potential donor star forms, 
the rate of interaction between this extremely compact binary with other single stars becomes low. 
However, this compact binary can continue to interact more frequently with other relatively wide 
binaries. As a result, binary--binary interactions are more frequent than binary--single interactions 
after the formation of a tight enough binary which will eventually become a MTBHB (Figure\ \ref{fig:CDF}). 
Furthermore, the compactness of the MTBHB progenitors also ensure that the potential donor star 
does not get exchanged out through a dynamical encounter. While binary--binary encounters are typically more common, the change in the semi-major axis of the MTBHB progenitor through binary--binary interactions is lower compared to the typical changes arising from binary--single encounters (Figure\ \ref{fig:deltaA}). This is because the tight MTBHB progenitors typically interact with the widest binaries available which cannot significantly alter the semimajor axis of the tight binary. 

Figure \ref{fig:CDF} demonstrates that dynamical interactions play a significant role in the formation and evolution of all MTBHBs. BHs, being more massive, sink 
in the cluster's potential. Thus the BHs in MTBHBs typically go through many strong encounters before the formation of the MTBHB. During this time, these encounters 
often lead to exchanges where the BHs can frequently acquire different companions. Eventually, the BH acquires a companion that would be the progenitor of the MTBHB. The configuration,
of course, needs to be compact enough that subsequent scattering interactions do not eject 
this companion. Following the creation of the MTBHB progenitor, subsequent interactions and/or tidal decay harden it to the point of RLO. Because of the dramatic dynamical history of MTBHBs, their 
production as well as their properties are entirely governed by the dynamical processes 
in the cluster and not by initial assumptions. Furthermore, the appearance of any {\em active} MTBHBs during any small time window in the cluster's life is extremely stochastic. 

\subsection{Properties of mass-transferring BH systems}

\begin{figure}[t!]
\plotone{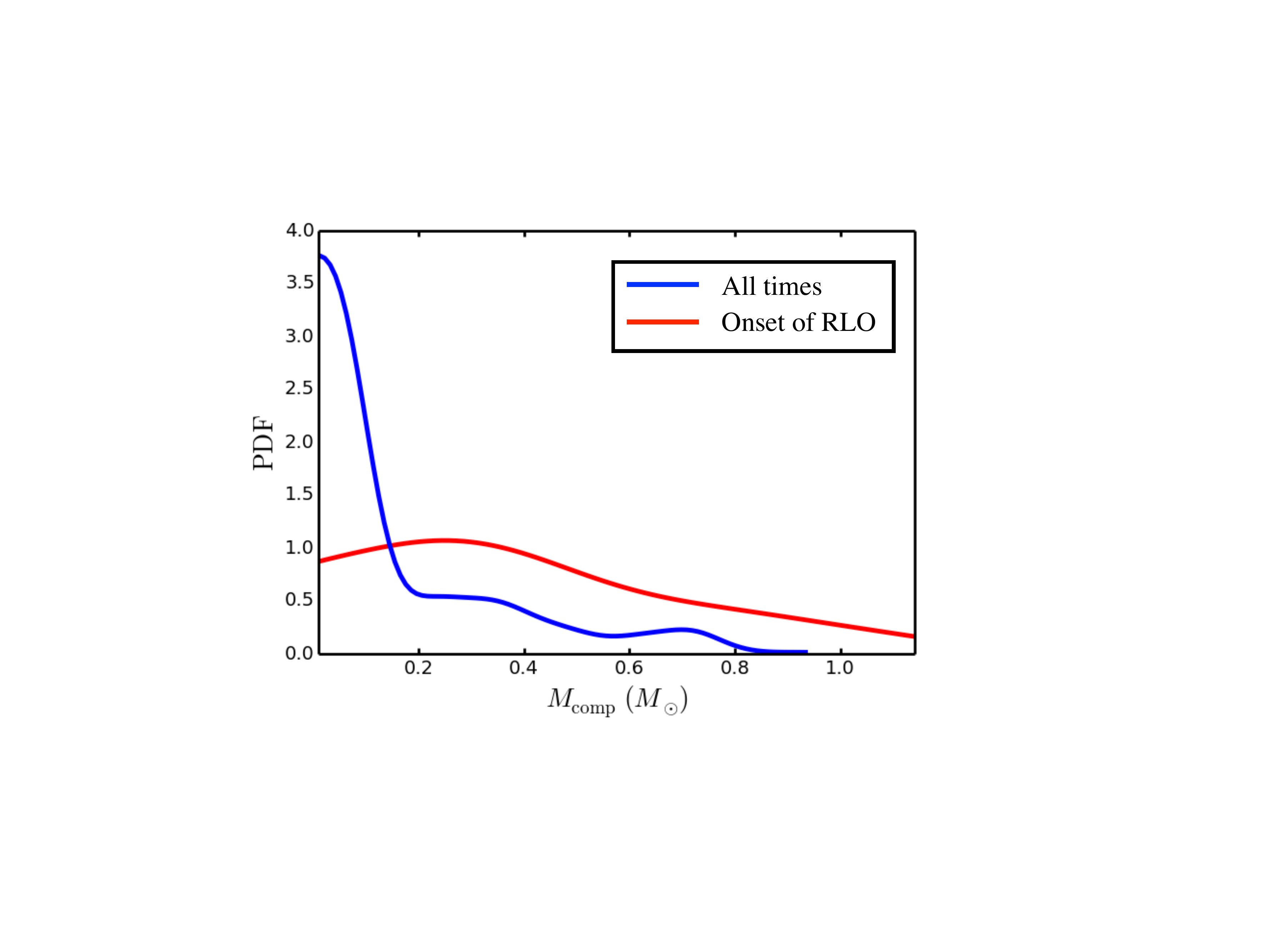}
\caption{\label{fig:Mcomp} \footnotesize The red curve shows the KDE of companion mass at the onset of mass transfer for all MTBHBs in our sample. The blue curve shows the KDE of the companion mass for \textit{all} snapshots in time for $t \geq 8$ Gyr.}
\end{figure}

Figure \ref{fig:Mcomp} shows the kernel-density estimate (KDE) of the mass of the companion star at the onset of RLO (red) and at every snapshot 
at late times ($t\geq8\,\gyr$; blue). Note that the $\Delta t$ between snapshots considered here is normalized to a constant value of 250 Myr to give equal weight to all models (with different snapshot frequencies). The red plot illustrates that companion masses at the onset of mass transfer are usually low. This is expected because all massive stars would have already collapsed into compact objects by the late times considered here ($t \geq 8$\ \gyr). 

The blue plot in Figure \ref{fig:Mcomp} shows that the observed MTBHBs in GCs are likely to have very low mass ($M \leq 0.1 M_{\odot}$). This is because the mass of the donor continuously decreases during mass transfer.

\begin{figure}[t!]
\plotone{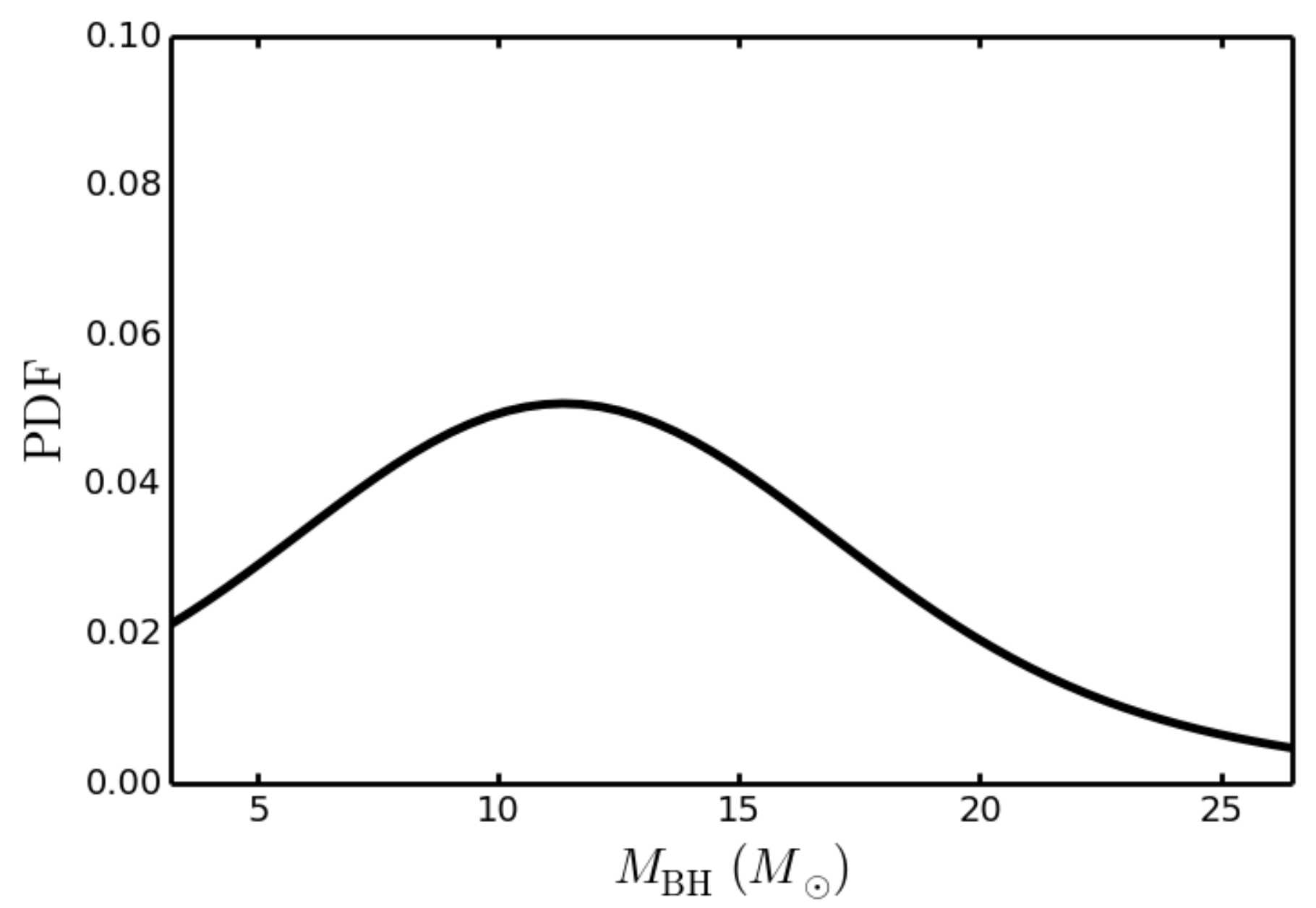}
\caption{\label{fig:Mbh} \footnotesize Kernel density estimate of black hole mass at the onset of mass transfer for all MTBHBs in our sample.}
\end{figure}

Figure \ref{fig:Mbh} shows the KDE of the mass of the BH at the onset of RLO.  As Figure \ref{fig:Mbh} shows, the BHs found in MTBHBs at late times are usually near the low end of the BH mass spectrum, with a median value of $\sim 11.8\,\msun$. This is consistent with our understanding of the dynamical evolution of BHs in GCs. The most massive BHs mass-segregate 
first, and are ejected first via mutual strong scattering encounters. 
Clusters as old as the GCs in the MW only retain
the relatively lower-mass BHs which can take part in 
exchange encounters with other non-BHs and form MTBHB progenitors. 


\subsection{Effect of triples}
\label{triples}
As the formation and evolution of triples has a negligible effect on the overall evolution 
of a GC, we do not consider the evolution of triples in \texttt{CMC}, for simplicity. However, 
because of the prevalence of binary-binary encounters that MTBHB progenitors 
go through before they are driven to RLO, triples do form in our simulations and may be relevant to the formation of MTBHBs. 

When the compact 
MTBHB progenitor is involved in a binary--binary encounter, the MTBHB progenitor often exchanges into a wider binary (replacing a single companion of the latter 
binary) creating a hierarchical triple. However, as expected from the interaction cross-section, 
typically the widest of the binaries, near the hard-soft boundary, interact with the progenitors of the MTBHBs. Since the binding energy of the outer binary is 
negligible compared to that of the inner binary (the MTBHB progenitor), breaking these triples 
does not typically significantly alter the fate of the inner binary. 

However, triple-mediated {\em secular} interactions may play a crucial role in hastening the 
RLO for the inner binary. For example, \citet[][]{Ivanova2010}  and \citet{Naoz2016} have noted that triples may play 
an important role in the formation of BH-XRBs in GCs. We now estimate an {\em upper limit} on 
the formation of MTBHBs via secular evolution of hierarchical triples in our simulations. 

In a dynamically stable triple, the gravitational interaction between the inner and outer binary drives periodic variations of the mutual inclination between the two orbits and the inner binary's eccentricity \citep{Lidov1962,Kozai1962}. These oscillations, known as Lidov-Kozai (LK) oscillations, may lead to close approach between the components of the inner binary, which may induce mass-transfer.




The characteristic timescale for LK oscillations is given by:
\begin{equation}
\label{eq:kozai}
T_{LK} \simeq P_{\rm{in}} \left(\frac{M_B}{M_S} \right) \left( \frac{a_{\rm{out}}}{a_{\rm{in}}} \right)^3 (1-e_{\rm{out}}^2)^{3/2}
\end{equation}
\citep{Holman1997}, where $P_{\rm{in}}$ is the orbital period of the inner binary, $M_B$ is the mass of the inner binary, $M_S$ is the mass of the outer star, $a_{\rm{out}}$ and $a_{\rm{in}}$ are the semi-major axes of the outer and inner binaries respectively, and $e_{\rm{out}}$ is the eccentricity of the outer binary. The 
characteristic time between subsequent dynamical interactions between the triple and other stars, incorporating the effects of gravitational focusing, can be expressed as:

\begin{equation}
\label{eq:dyntime}
t_{\rm{dyn}} = \frac{1}{\pi} n^{-1}  a_{\rm{out}}^{-2}\,\sigma_v^{-1} \bigg[ 1 + \frac{G\,\big(M_{\rm{trip}}+\langle m \rangle \big)}{2\,a_o\,\sigma_v^2} \bigg]^{-1}.
\end{equation}
Here $n$ and $\sigma_v$ are the stellar density and velocity dispersion, respectively, in the region occupied by the triple,  $M_{\rm{trip}}$ is the total mass of the triple ($M_B + M_S$), and $\langle m \rangle$ is the mass of an average star in the cluster.

For any triple with a BH---non-BH inner binary that satisfies $T_{LK}\leq t_{\rm{dyn}}$, LK oscillations may drive the system to RLO before the next dynamical encounter can break or significantly alter the outer orbit.  In all of our models, we find a total of 58 BHs which (1) had gone through a phase where it was part of the inner binary of a dynamically formed triple which satisfied the above criteria at any point of time and (2) are retained within their host clusters at $t \geq 8$ Gyr. The relevant parameters for each of these 58 triples are shown in Table \ref{table:3}. In order to determine which of these 58 systems could be driven to RLO due to LK oscillations during this triple phase, we use the Octupole-Level Secular Perturbation Equations (\texttt{OSPE}) package \citep{Naoz2011, Naoz2013} to integrate each triple over $t_{\rm{dyn}}$. If at any point during this integration, the eccentricity of the inner binary, $e_{\rm{in}}$ is driven to a value satisfying

\begin{equation}
\label{eq:tripleMT}
R_{\rm{comp}} \geq (1-e_{\rm{in}})\,R_L,
\end{equation}
the system will likely undergo mass-transfer. Here $R_{\rm{comp}}$ is the stellar radius of the BH's companion star in the inner binary and $R_L$ is the Roche lobe of the companion, given by Equation (\ref{eq:RL}).

\startlongtable
\begin{deluxetable*}{c|c|c|ccc|cc|cc|c|c|c}
\tabletypesize{\scriptsize}
\tablecolumns{14}
\tablewidth{0pt}
\tablecaption{Triple Properties. \label{table:3}}
\tablehead{
	\colhead{Binary} &
    \colhead{$t_{\rm{triple}}$} &
    \colhead{Donor Type} &
    \colhead{$M_{\rm{BH}}$} &
    \colhead{$M_{\rm{comp}}$} &
    \colhead{$M_3$} &
    \colhead{$a_{\rm{out}}$} &
    \colhead{$a_{\rm{in}}$} &
    \colhead{$e_{\rm{out}}$} &
    \colhead{$e_{\rm{in}}$} &
    \colhead{$R_{\rm{comp}}$} &
    \colhead{$t_{\rm{dyn}}$} &
    \colhead{$\rm{MT}_{\rm{frac}}$} \\
    \cline{4-6}\cline{7-8}
    \colhead{} & 
    \colhead{($\gyr$)} &
    \colhead{} &
    \multicolumn{3}{c}{($\msun$)} &
    \multicolumn{2}{c}{(AU)} &
    \colhead{} &
    \colhead{} &
    \colhead{($R_\odot$)} &
    \colhead{(Myr)}
}
\startdata
1 & 8.42 & MS & 13.44 & 0.72 & 1.3 & 8.8 & 0.51 & 0.76 & 0.4 & 0.706 & 8.4 & 0.2 \\
2 & 8.62 & WD & 13.44 & 1.05 & 1.37 & 34.5 & 1.08 & 0.76 & 0.62 & 0.008 & 1.7 & 0 \\
3 & 7.42 & WD & 14.46 & 1.36 & 1.38 & 462.7 & 96.64 & 0.42 & 0.72 & 0.002 & 0.3 & 0 \\
4 & 4.76 & MS & 16.12 & 0.16 & 16.83 & 190.5 & 21.28 & 0.49 & 0.59 & 0.186 & 2.3 & 0.3 \\
5 & 9.26 & MS & 7.54 & 0.41 & 18.44 & 311.0 & 1.43 & 0.76 & 0.49 & 0.367 & 0.8 & 0 \\
6 & 6.39 & WD & 7.6 & 0.45 & 1.02 & 11.9 & 0.16 & 0.3 & 0.46 & 0.008 & 192.5 & 0 \\
7 & 10.72 & MS & 14.66 & 0.43 & 16.63 & 55.1 & 0.46 & 0.62 & 0.65 & 0.39 & 8.5 & 0 \\
8 & 11.51 & WD & 14.74 & 8.23 & 1.28 & 148.8 & 20.4 & 0.5 & 0.98 & 0.004 & 0.4 & 0.1 \\
9 & 0.74 & MS & 14.11 & 0.21 & 25.31 & 167.6 & 2.37 & 0.82 & 0.92 & 0.228 & 1.5 & 0.3 \\
10 & 13.57 & WD & 7.98 & 0.95 & 0.84 & 17.7 & 0.47 & 0.77 & 1.0 & 0.009 & 4.5 & 0 \\
11 & 14.03 & MS & 8.93 & 0.23 & 1.17 & 2.2 & 0.33 & 0.56 & 0.1 & 0.242 & 9.3 & 0 \\
12 & 8.8 & MS & 13.52 & 0.71 & 1.31 & 485.6 & 142.89 & 0.23 & 0.82 & 0.697 & 0.7 & 0.1 \\
13 & 9.46 & MS & 18.15 & 0.43 & 0.71 & 209.4 & 12.91 & 0.65 & 0.58 & 0.707 & 0.8 & 0.1 \\
14 & 13.86 & WD & 18.17 & 1.31 & 1.2 & 2.3 & 0.27 & 0.61 & 0.98 & 0.003 & 17.3 & 0 \\
15 & 11.27 & MS & 8.79 & 0.27 & 13.24 & 22.5 & 0.05 & 0.52 & 0.53 & 0.272 & 18.0 & 0 \\
16 & 8.58 & MS & 13.92 & 0.06 & 14.66 & 25.6 & 0.01 & 0.86 & 0.0 & 0.173 & 19.9 & 1.0 \\
17 & 8.65 & MS & 14.28 & 0.21 & 0.16 & 141.1 & 9.19 & 0.76 & 0.88 & 0.224 & 11.9 & 0.4 \\
18 & 10.25 & MS & 13.49 & 0.52 & 16.27 & 180.1 & 7.57 & 0.02 & 0.44 & 0.464 & 1.0 & 0.1 \\
19 & 8.49 & MS & 7.44 & 0.3 & 13.72 & 31.5 & 1.16 & 0.58 & 0.3 & 0.291 & 26.0 & 0.3 \\
20 & 11.0 & WD & 3.23 & 1.24 & 0.95 & 4.9 & 0.19 & 0.68 & 0.21 & 0.004 & 16.7 & 0 \\
21 & 6.33 & MS & 9.53 & 6.44 & 13.52 & 37.8 & 3.54 & 0.32 & 0.59 & 0.8 & 14.3 & 0.1 \\
22 & 10.36 & MS & 11.26 & 0.5 & 11.9 & 24.4 & 0.05 & 0.9 & 0.55 & 0.45 & 14.5 & 0 \\
23 & 10.61 & MS & 10.1 & 0.47 & 12.61 & 133.6 & 0.35 & 0.8 & 0.77 & 0.427 & 1.4 & 0 \\
24 & 9.25 & MS & 8.68 & 0.21 & 0.48 & 21.8 & 1.0 & 0.82 & 0.96 & 0.23 & 31.5 & 0.8 \\
25 & 8.75 & MS & 13.73 & 0.87 & 13.8 & 175.0 & 0.49 & 0.93 & 0.83 & 1.257 & 0.3 & 0 \\
26 & 9.27 & WD & 10.79 & 0.24 & 11.99 & 36.1 & 0.11 & 0.91 & 0.98 & 0.02 & 2.6 & 0 \\
27 & 3.81 & G & 17.46 & 1.14 & 13.86 & 12.2 & 0.15 & 0.31 & 0.0 & 6.223 & 258.9 & 1.0 \\
28 & 10.63 & MS & 12.33 & 0.77 & 1.2 & 6.1 & 0.12 & 0.33 & 0.48 & 0.875 & 5.0 & 0 \\
29 & 10.9 & MS & 12.33 & 0.77 & 0.49 & 4.5 & 0.11 & 0.85 & 0.61 & 0.885 & 5.9 & 0.3 \\
30 & 11.62 & WD & 5.6 & 1.33 & 0.72 & 5.0 & 0.19 & 0.79 & 0.91 & 0.002 & 12.7 & 0 \\
31 & 9.49 & MS & 7.06 & 0.52 & 16.53 & 26.5 & 0.04 & 0.26 & 0.52 & 0.468 & 40.2 & 0 \\
32 & 10.58 & MS & 13.88 & 9.54 & 11.87 & 1910.3 & 38.5 & 0.8 & 0.7 & 0.796 & 29.0 & 0 \\
33 & 11.37 & MS & 11.27 & 0.16 & 12.71 & 83.8 & 1.68 & 0.66 & 0.89 & 0.179 & 36.8 & 0.1 \\
34 & 6.99 & MS & 13.0 & 0.33 & 24.76 & 129.0 & 0.94 & 0.92 & 0.58 & 0.312 & 0.4 & 0.2 \\
35 & 7.22 & WD & 4.64 & 0.8 & 0.56 & 8.1 & 0.19 & 0.75 & 0.55 & 0.013 & 55.4 & 0 \\
36 & 9.55 & WD & 3.74 & 1.27 & 7.02 & 144.3 & 0.52 & 0.88 & 0.99 & 0.004 & 0.3 & 0 \\
37 & 10.89 & MS & 7.97 & 0.7 & 1.3 & 3.4 & 0.29 & 0.31 & 0.87 & 0.701 & 36.5 & 0.2 \\
38 & 7.47 & MS & 13.72 & 0.63 & 9.9 & 130.2 & 1.49 & 0.5 & 0.71 & 0.578 & 8.2 & 0 \\
39 & 4.66 & WD & 11.23 & 0.98 & 1.2 & 22.2 & 1.57 & 0.55 & 0.16 & 0.005 & 16.8 & 0.1 \\
40 & 10.67 & MS & 11.78 & 0.42 & 0.41 & 8.5 & 0.4 & 0.69 & 0.51 & 0.384 & 16.3 & 0.3 \\
41 & 11.49 & WD & 4.06 & 1.15 & 0.67 & 12.6 & 0.24 & 0.8 & 0.9 & 0.006 & 10.4 & 0 \\
42 & 5.41 & MS & 10.55 & 0.55 & 0.71 & 11.6 & 0.24 & 0.87 & 0.55 & 0.495 & 37.2 & 0.9 \\
43 & 8.08 & MS & 4.88 & 0.75 & 1.75 & 3.0 & 0.26 & 0.56 & 0.52 & 1.273 & 126.6 & 0.2 \\
44 & 11.39 & MS & 7.17 & 0.83 & 11.54 & 170.6 & 0.91 & 0.77 & 0.57 & 1.414 & 1.8 & 0.1 \\
45 & 8.7 & WD & 14.0 & 1.3 & 19.21 & 247.0 & 37.54 & 0.38 & 0.7 & 0.003 & 1.9 & 0 \\
46 & 6.85 & MS & 8.06 & 0.73 & 0.83 & 5.0 & 1.32 & 0.16 & 0.35 & 0.9 & 128.5 & 0 \\
47 & 10.03 & WD & 6.57 & 1.25 & 11.81 & 134.1 & 1.0 & 0.94 & 0.68 & 0.004 & 0.7 & 0 \\
48 & 10.03 & WD & 5.96 & 1.28 & 6.57 & 4.6 & 1.09 & 0.18 & 0.8 & 0.004 & 17.0 & 0 \\
49 & 11.25 & WD & 3.94 & 1.33 & 1.18 & 4.2 & 0.4 & 0.28 & 0.99 & 0.003 & 8.9 & 0 \\
50 & 11.46 & WD & 3.92 & 1.32 & 4.33 & 27.5 & 0.29 & 0.65 & 0.34 & 0.016 & 0.3 & 0 \\
51 & 0.44 & MS & 21.29 & 0.14 & 29.82 & 24.8 & 0.04 & 0.55 & 0.13 & 0.159 & 4.2 & 0 \\
52 & 0.86 & MS & 26.48 & 0.25 & 26.07 & 283.7 & 8.03 & 0.43 & 0.96 & 0.257 & 26.7 & 0.1 \\
53 & 5.38 & MS & 14.11 & 0.5 & 13.75 & 3.2 & 0.14 & 0.16 & 0.82 & 0.438 & 113.4 & 0.2 \\
54 & 1.53 & MS & 18.44 & 0.11 & 22.36 & 9.0 & 0.07 & 0.31 & 0.67 & 0.136 & 86.3 & 0 \\
55 & 10.97 & MS & 7.91 & 0.72 & 0.86 & 1.2 & 0.13 & 0.66 & 0.44 & 0.01 & 34.0 & 0 \\
56 & 1.89 & MS & 13.77 & 0.18 & 17.83 & 26.1 & 0.04 & 0.76 & 0.13 & 0.203 & 11.0 & 0 \\
57 & 5.06 & MS & 15.95 & 0.11 & 18.39 & 15.5 & 0.23 & 0.89 & 0.48 & 0.137 & 190.3 & 0.2 \\
58 & 2.3 & MS & 18.35 & 0.29 & 24.37 & 9.4 & 0.08 & 0.91 & 0.44 & 0.283 & 23.2 & 0.2 \\
\enddata
\tablecomments{Orbital parameters for all triple systems that were integrated using \texttt{OSPE}. $t_{\rm{triple}}$ denotes time of 
formation of the triple. $\rm{MT}_{\rm{frac}}$ gives the fraction of \texttt{OSPE} integrations which resulted in Roche-lobe overflow.
}
\end{deluxetable*}

\texttt{OSPE} takes as input the masses of the three bodies, the semi-major axis and eccentricity of the inner and outer binaries, the mutual inclination, $i$, of the inner and outer binaries, and the arguments of pericenter, $g_1$ and $g_2$, for the two non-central (less-massive) bodies. Because we lack the values for the latter three parameters, we sample the $\cos i$ uniformly from $[-1,1]$ and sample $g_1$ and $g_2$ uniformly from $[0,2 \pi]$.

We model 10 independent integrations for each of our 58 triples in order to statistically determine which of these systems are likely to be driven to mass-transfer. Of the 580 total triples integrated, 79 systems ($\sim 14 \%$) are driven to RLO. The final column of Table \ref{table:3}, $\rm{MT}_{\rm{frac}}$,  gives the fraction of integrations which resulted in RLO for each triple.

26 of the 58 total triples are driven to RLO in at least one \texttt{OSPE} integration. Of these 26 systems, 23 have main sequence donors, 1 has a giant donor, and 2 have white dwarf donors. These values are consistent with the relative values of each donor type for the 17 MTBHBs produced through the binary-mediated/dynamics channel (Table \ref{table:2}).

If a system is driven to mass-transfer by LK-oscillations, it is unclear how long the system will remain in a mass-transferring state. In particular, if a system begins mass-transfer through a triple-mediated event at a time earlier than $\sim 8$ \gyr, it is not certain that this system will still be mass-transferring at the present age of the cluster. Of the 79 systems that are driven to RLO during the \texttt{OSPE} integrations, 43 systems (or $\sim 7 \%$ of the 580 total) are driven to mass-transfer after $t = 8$ \gyr.

Applying the $7\%$ and $14\%$ cuts to the 58 triple systems identified, we conclude that in our 50 GC models, 4--8 MTBHBs may exist which were formed through the LK-oscillation/triple-mediated channel. We note that these numbers are small relative to the number of systems formed through the binary-mediated/dynamics channel (17 systems). 

Note that our treatment of triples does not consider the effect of tidal interactions on the inner BH--non-BH binary. It is possible that an eccentricity boost from LK-oscillations may initiate tidal decay within the inner BH--non-BH binary of the triple, and ultimately drive that system to Roche-lobe overflow before the (eccentric) Roche-lobe overflow condition (Equation \ref{eq:tripleMT}) is reached. This effect could potentially slightly raise the upper limit of the population of the triple-mediated MTBHBs.

\section{Numbers of MTBHBs vs Retained Black Holes}
\label{sec:rates}

We now investigate what finding a MTBHB in a cluster indicates about the overall retained 
BH population in that cluster. 
To  explore the relation of the MTBHBs identified in our models with the total number of BHs retained ($N_{\rm{BH}}^{\rm{tot}}$), we group together those MTBHBs formed through both the standard binary-mediated/dynamics channel (binaries in Table \ref{table:2}) and the LK-oscillations/triple-mediated channel (binaries in Table \ref{table:3}).

Because the \texttt{OSPE} triple integrations are performed outside of the \texttt{CMC} simulations, we cannot determine how long a system which is driven to RLO through LK-oscillations will continue to mass-transfer. Instead, we consider three possibilities for the contribution of triples to the total number of MTBHBs in our models. For Case\ 1, we assume that any triple which is driven to mass-transfer in at least one of the ten \texttt{OSPE} integrations, independent of when RLO begins, is a MTBHB and will remain a MTBHB from the time the triple is formed until the present day (assuming the BH is not ejected from its host cluster). Case\ 2 is identical to the first case, with the additional stipulation that the triple must have formed after $t=8$ Gyr. For Case\ 3, we neglect the contribution from the triple-mediated channel entirely. 

\begin{figure}[t!]
\plotone{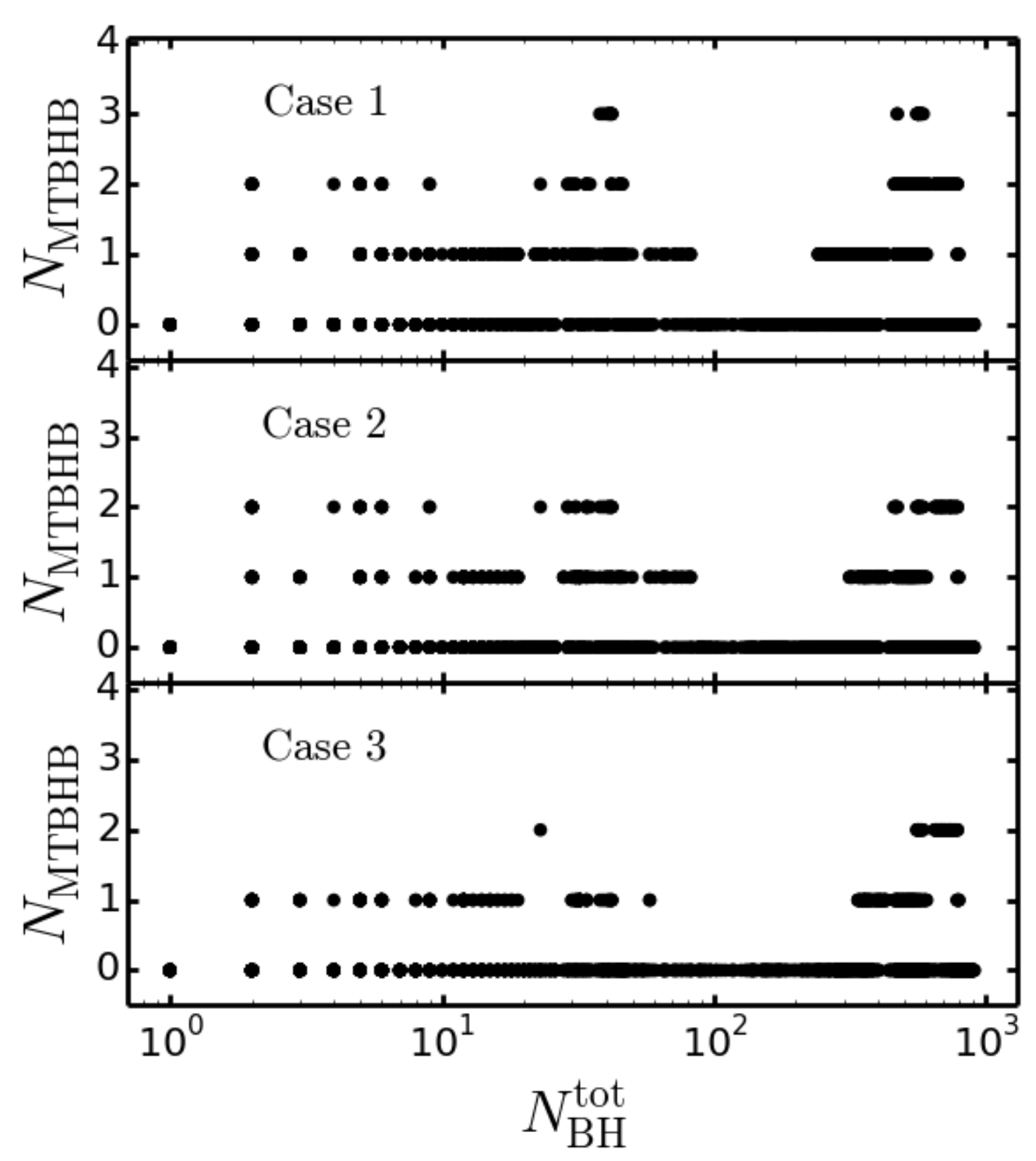}
\caption{\label{fig:mass-transferring BH binariesvsBH} \footnotesize Number of MTBHBs in each model versus total number of BHs retained in each model for all snapshots in time with $t \geq 8$ Gyrs. Case\ 1 includes the contribution from any triple driven to mass-transfer in at least one of the ten \texttt{OSPE} integrations. Case\ 2 is identical to the first case, with the additional stipulation that the triple must have formed after $t=8$ Gyr. Case\ 3 neglects the contribution from the triple-mediated channel entirely. }
\end{figure}

Figure \ref{fig:mass-transferring BH binariesvsBH} shows the relation between the number of MTBHBs, $N_{\rm{MTBHB}}$, and the total number of BHs, $N_{\rm{BH}}^{\rm{tot}}$, within each of our models for the three cases described above. Here, we simply count the total number of BHs retained within each model in each snapshot in time. All our snapshots are roughly equidistant in time, so we 
treat each snapshot at $t \geq 8$ Gyr as a single observed cluster where MTBHBs may have been found. 

As Figure \ref{fig:mass-transferring BH binariesvsBH} shows, the number of MTBHBs is uncorrelated with the total number of BHs the cluster contains, regardless of the details of the contribution from triple-mediated formation channels. The Spearman correlation coefficient between the $N_{\rm{MTBHB}}$ and $N_{\rm{BH}}^{\rm{tot}}$ is 0.13, 0.05, and 0.03 for Cases 1, 2, and 3, respectively. 
This result is in agreement with the result shown in \citet{Chatterjee2017a}, which used the total number of BH--non-BH (BH--nBH) binaries as a proxy to the upper limit on the total number of MTBHBs and demonstrated a similar lack of correlation. 

The lack of correlation between $N_{\rm{MTBHB}}$ and $N_{\rm{BH}}^{\rm{tot}}$ shown in Figure \ref{fig:mass-transferring BH binariesvsBH} can be understood by a close inspection 
of the rate of dynamical formation of BH-nBH binaries.

Because of mass segregation, the stars and binaries in the cluster are segregated at cluster-centric 
distances based on their mass. 
MTBHBs can dynamically form only in a region  
of the cluster where BHs mix with potential mass-transferring companion stars, such as main sequence (MS) stars, giants, and white dwarfs. We define this mixing zone as a radial shell whose inner radius is determined by the radial position of the innermost MS star, giant, or white dwarf and whose outer radius is determined by the radial position of the cluster's outermost BH that lies within the observed core radius of the cluster. Note that as a result 
of recoil from scattering encounters some BHs can remain outside the core radius before 
it sinks to the core again due to mass segregation. We ignore them because, 
the binary-mediated interaction cross-section involving BHs and non-BHs 
is expected to be dominated by only the BHs within the core where the stellar 
density is relatively high. Additionally, inside the core the density is roughly constant and, as a result, the 
calculation of interaction rates becomes simpler. 

Single BHs can interact with non-BH--non-BH (nBH--nBH) binaries to form new BH--nBH binaries via 
exchange. On the other hand, single BHs can also interact with BH--nBH binaries to destroy the BH--nBH binary and instead create a BH--BH binary via exchange.   
The formation rate, $\Gamma$, of potential MTBHBs within this mixing zone can be expressed as
\begin{equation}
\label{form-o}
\Gamma_{\rm{form}} = n_{\rm{nBH-nBH}} \, \Sigma \, \sigma_v \, N_{\rm{BH}} B_{\rm{ex}}
\end{equation}
where $n_{\rm{nBH-nBH}}$ is the number density of binaries in which both components are non-BHs in the mixing zone, $\Sigma$ is the cross section for interaction between nBH--nBH binaries and other BHs in the mixing zone, $\sigma_v$ is the relative velocity dispersion of nBH--nBH binaries and BHs, and $N_{\rm{BH}}$ is the total number of BHs in the mixing zone. $B_{\rm{ex}}$ is the exchange rate for interactions between nBH--nBH binaries and BHs.

Of course, the larger the semi-major 
axis, $a$, of the nBH--nBH binaries, the higher the interaction rate. As a result, binaries near the hard-soft boundary given 
by
\begin{equation}
\label{eq:hardsoft}
\frac{G \, \langle m_{\rm{nBH}} \rangle^2}{2a_{\rm{h-s}}}  = \frac{1}{2}\langle m \rangle \, \sigma_v^2,
\end{equation}
where, $a_{\rm{h-s}}$ is the hard-soft boundary, are the ones that interact most often. Here $\langle m_{\rm{nBH}} \rangle$ is the average mass of the components of nBH--nBH binaries in the mixing zone, and $\langle m \rangle$ is the average mass of all stars in the mixing zone. Assuming that the overall interaction rate 
is dominated by the binaries with a semi-major axis $a \sim a_{\rm{h-s}}$ and that $\Sigma \sim a_{\rm{h-s}}^2$
allows us to re-write Equation (\ref{form-o}) as:

\begin{equation}
\label{form}
\Gamma_{\rm{form}} \propto n_{\rm{nBH-nBH}} \, \frac{ \langle m_{\rm{nBH}} \rangle ^4}{\langle m \rangle ^2} \frac{N_{\rm{BH}}} {\sigma_v^3} 
\end{equation}
where we have assumed $B_{\rm{ex}}$ to be independent of $N^{\rm{tot}}_{\rm{BH}}$, a reasonable assumption.
%
Similarly, the rate of destruction of BH--nBH binaries can be expressed as
\begin{equation}
\label{destruct}
\Gamma_{\rm{destruct}} \propto n_{\rm{BH-nBH}} \, \frac{ \langle m_{\rm{nBH}} \rangle ^2 \langle m_{\rm{BH}} \rangle ^2}{\langle m \rangle ^2} \frac{N_{\rm{BH}}} {\sigma_v^3} 
\end{equation}
where $n_{\rm{BH-nBH}}$ is the number density of BH--nBH binaries in the mixing zone. 
Combining Equations (\ref{form}) and (\ref{destruct}) we obtain the net rate of 
formation 

\begin{multline}
\label{eq:rate}
\Gamma_{\rm{total}} \propto  \frac{ \langle m_{\rm{nBH}} \rangle^2}{\langle m \rangle ^2} \frac{N_{\rm{BH}}}{\sigma_v^3} \times \\
\Bigg[  n_{\rm{nBH-nBH}} \langle m_{\rm{nBH}} \rangle ^2  -  n_{\rm{BH-nBH}} \langle m_{\rm{BH}} \rangle ^2\Bigg]
\end{multline}

On the basis of the lack correlation between $N_{\rm{MTBHB}}$ and $N_{\rm{BH}}^{\rm{tot}}$ shown in Figure \ref{fig:mass-transferring BH binariesvsBH}, one would expect 
that the net formation rate of MTBHBs ($\Gamma_{\rm{total}}$) 
within a cluster is independent of the total number of BHs the cluster contains.

\begin{figure}[t!]
\plotone{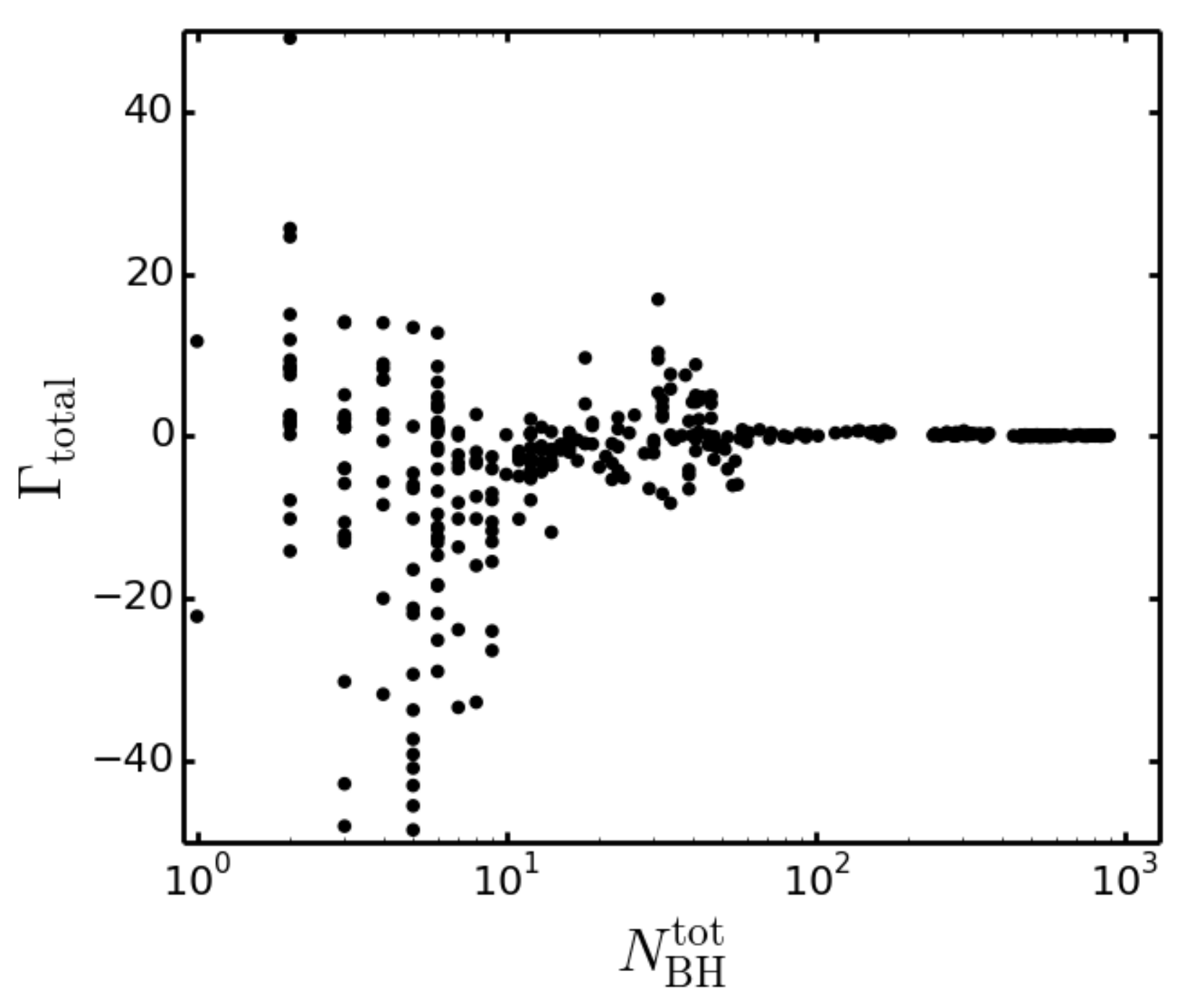}
\caption{\label{fig:crosssection} \footnotesize The total formation rate of BH--nBH binaries in the mixing zone, $\Gamma_{\rm{total}}$ (Eq.\ \ref{eq:rate}), plotted versus total number of BHs, $N^{\rm{tot}}_{\rm{BH}}$. Each black circle represents a different snapshot in time for each our GC models. 
}
\end{figure}

Figure \ref{fig:crosssection} shows $\Gamma_{\rm{total}}$ plotted against $N^{\rm{tot}}_{\rm{BH}}$ for all snapshots in time with $t \geq 8$ Gyr for all GC models. Since for our purposes we are only interested in the 
functional form, we ignore the constants (such as $G$, $\pi$, $B_{\rm{ex}}$, etc.). All masses are in $\msun$, number densities are in $\pc^{-3}$, and velocities are in $\kms$. The spread in values for low $N_{\rm{BH}}^{\rm{tot}}$ is a consequence of the stochastic nature of the process magnified by the low numbers of BHs in the mixing zone, as well as varying initial cluster parameters (e.g. concentration, $N$, $r_v$) between our different models. The width of the spread can be viewed as the error on the $\Gamma_{\rm{total}}$ calculation. Clearly, the net rate of formation of potential MTBHBs does not depend on the total number of retained BHs for $N^{\rm{tot}}_{\rm{BH}}$ spanning 4 orders of magnitude.

Physically, this result can be explained as follows: When large numbers of BHs are present in a GC, the core of the cluster is dominated by the BHs due to mass segregation. 
Therefore while $N_{\rm{BH}}^{\rm{tot}}$ is large, the very central high-density regions are dominated primarily by single BHs, and the lower-mass stars (potential donors) are driven out of the central regions. Also, while a large number of BHs are still retained in the cluster, the energy production 
due to BH-dynamics keeps the stellar density in the mixing zone, typically further away from the center, low. 
Thus, the internal dynamics of such a BH-dominated center makes it difficult for these BHs to dynamically 
acquire a non-BH companion. In this regime the rate limiting factor is $n_{\rm{nBH-nBH}}$.  
Thus, in spite of a large number of retained BHs in the cluster, the formation rate of potential MTBHBs remains low.

Only after a cluster is sufficiently depleted of its BHs, can the BHs significantly mix with 
other stars in the cluster. In this regime of low $N_{\rm{BH}}^{\rm{tot}}$, the outer radius of the mixing zone moves closer to the center, and density in the mixing zone 
increases. All of these increase the dynamical formation rate of BH--nBH binaries per 
BH. However, at this stage, total BH--nBH formation rate is limited by the number 
of BHs present in the cluster. 
Thus, the formation efficiency of MTBHBs remains self-regulated and largely independent of the total number of retained BHs in a cluster at all late times. 

\section{Ejected Black Hole Binaries}
\label{sec:ejected}

Throughout a cluster's evolution, a large number of BHs are ejected from the cluster through both dynamical encounters and supernova kicks. If these BHs are ejected as members of BH-nBH binaries, these systems may eventually become mass-transferring, although they would 
not be identified as cluster members. In this manner, GCs may contribute to the population of low-mass X-ray binaries (LMXBs) with BH accretors in the galactic halo \citep[e.g.,][]{Giesler2017}. The present-day location of these ejected systems within the galactic halo depends upon the time of ejection as well as the ejection velocity.

\begin{figure}[t!]
\plotone{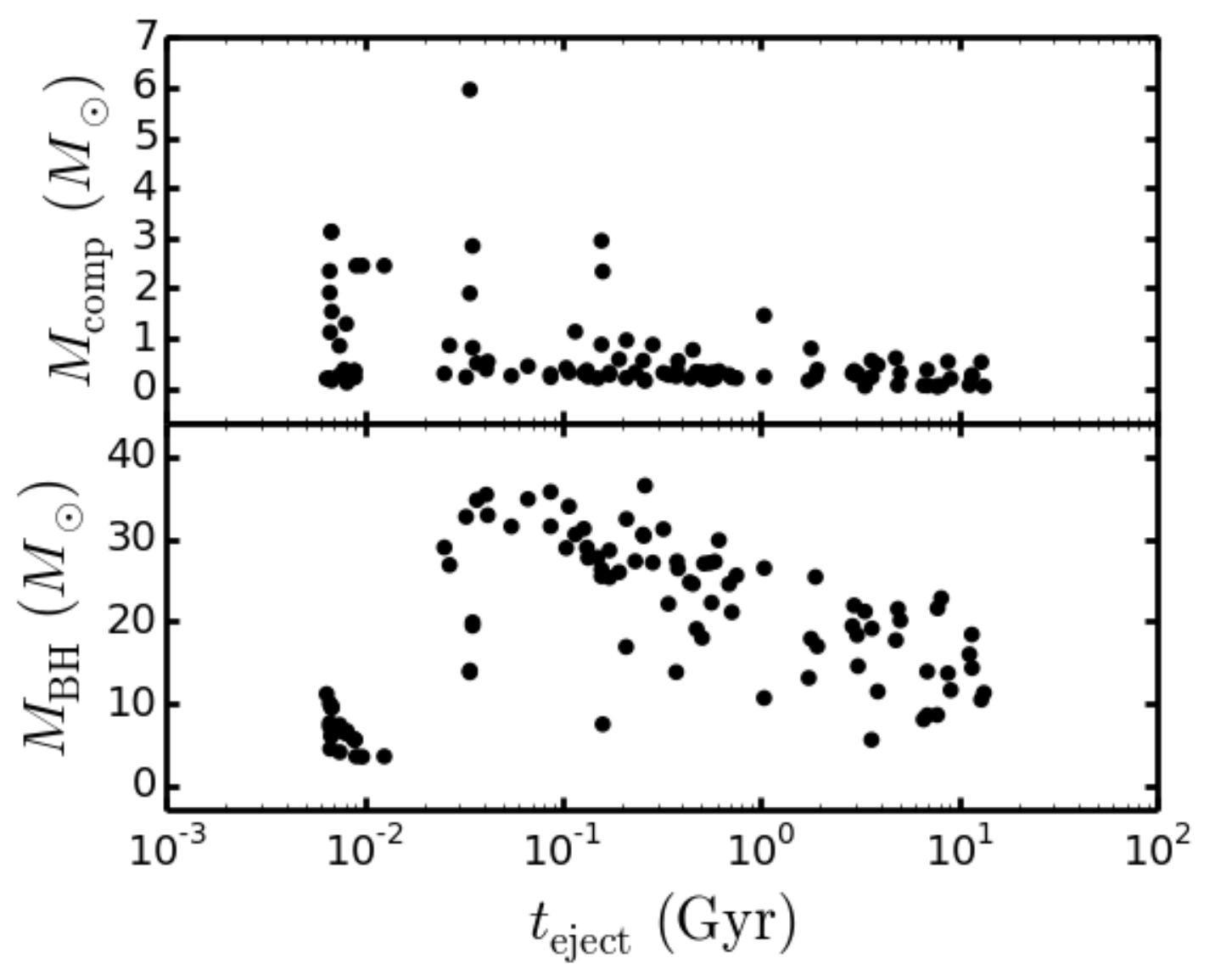}
\caption{\label{fig:XRBejected} \footnotesize Masses of all companion stars (top panel) and BHs (bottom panel) versus time of ejection for all  ejected BH-nBH binaries that eventually become mass-transferring with $8\leq t/\gyr\leq13$.}
\end{figure}

For all BH-nBH binaries ejected from our cluster models, we  determine whether or not each system become mass-transferring by evolving the system as an isolated binary in \texttt{BSE} with initial properties similar to the properties of the binary at 
the time of its escape from the cluster. Figure \ref{fig:XRBejected} shows the masses 
of BHs (bottom panel) and their non-BH companions (top panel) for ejected BH-nBH binaries that eventually become mass transferring within the time range of 8-13 Gyr. 

As the top panel of Figure \ref{fig:XRBejected} illustrates, the companions in these systems tend to be low-mass stars, in line with the mass-transferring systems which are retained within the clusters. The downward trend of BH masses versus time of ejection seen in the bottom panel is reflective of our understanding of the dynamical processing of BHs in GCs. 
The more massive BHs are ejected from the cluster earlier than the less massive BHs since the more massive 
BHs mass segregate first and take part in dynamical encounters earlier than the lower-mass counterparts. As a result, the  average mass of BHs retained within a cluster decreases over time \citep{Morscher2013, Morscher2015}. Thus, the BHs in BH--nBH binaries, and MTBHBs created and ejected from the cluster at 
later times tend to have lower masses compared to those created and ejected earlier. 

A more thorough analysis of these ejected BH-XRB candidates, which explores the orbital parameters of these systems after they escape the cluster and their present day locations 
in the galactic halo will be presented in a later paper.

\section{Conclusion} \label{sec:conclusion}
We have studied the formation of MTBHBs in detailed cluster models with properties typical of the GCs found in the MW. Using models with a broad range of initial cluster parameters we found that {\em all} MTBHBs at late times ($t\geq8\,\gyr$) in a cluster are created dynamically and that none of the accreting BHs hold onto their primordial companions (Section\ \ref{sec:results}). The accreting BHs suffer a median of $\sim 30$ strong scattering events (Figure\ \ref{fig:CDF}). 
There are two main channels for the dynamical formation of MTBHBs: 
(1) formation via a series of binary-mediated scattering encounters that act in conjunction with orbital tidal decay to harden the system to Roche-lobe overflow, and (2) formation of a triple system with a BH-non-BH inner binary which is driven to mass transfer through Lidov-Kozai oscillations. 
We further found that the binary-mediated channel, where either no triple was ever 
formed or triple formation was not dynamically important in the evolution of the 
system, contribute more 
to the overall production of MTBHBs at 
late times, 
although the potential contribution 
from the triple-mediated channel may be significant (between $19$--$32\%$ of all MTBHB systems).

At late times, the MTBHBs typically have low-mass donors (Figure\ \ref{fig:Mcomp}). The BHs also are on the low-mass side of all BHs formed in the cluster (Figures\ \ref{fig:Mbh}). 

We also show that the number of MTBHBs in a cluster is independent of the total 
number of BHs retained in the cluster for BH numbers spanning 4 orders of magnitude.  
This lack of correlation can be understood by examining the BH--non-BH binary formation rate (Section\ \ref{sec:rates}). Specifically, the dynamical formation of MTBHBs in a cluster is a self-regulated process limited by a complex competition between the number of BHs, and the number density of non-BH--non-BH binaries in the zone of a cluster where BHs are mixed with the non-BHs. This competition keeps the number of MTBHBs formed in a cluster at any given time independent of the total number of BHs retained in the cluster at that time. 


\acknowledgments
We thank Smadar Naoz for kindly sharing the triple-evolution code \texttt{OSPE}. We also thank the referee for their useful suggestions.
This work was supported by NASA ATP Grant NNX14AP92G 
and NSF Grants AST--1312945 and AST--1716762. K.K. acknowledges support by the National Science Foundation Graduate Research Fellowship Program under Grant DGE--1324585.
S.C. acknowledges support from
CIERA, and from NASA Chandra Award TM5--16004X/NAS8--03060, issued by the Chandra X-ray Observatory Center
(operated by the Smithsonian Astrophysical Observatory
for and on behalf of NASA under contract NAS8-03060), 
and Hubble Space Telescope Archival research 
grant HST-AR-14555.001-A (from the Space Telescope 
Science Institute, which is operated by the Association of Universities for Research in Astronomy, Incorporated, under NASA contract NAS5-26555).

\software{\texttt{CMC} \citep{Joshi2000,Joshi2001,Fregeau2003, Fregeau2007, Chatterjee2010,Chatterjee2013,Umbreit2012,Morscher2013,Rodriguez2016b}, \texttt{Fewbody} \citep{Fregeau2004}, \texttt{BSE} \citep{Hurley2002}, \texttt{OSPE} \citep{Naoz2011, Naoz2013}}

\clearpage


\clearpage


\end{document}